\documentclass[traditabstract,twocolumn]{aa} 
\usepackage{color}

\usepackage[normalem]{ulem}
\usepackage{graphicx}
\usepackage{txfonts}
\usepackage{subfigure}
\usepackage{longtable,lscape}
\usepackage{natbib}
\bibpunct{(}{)}{;}{a}{}{,} 
\begin{document}

\newpage
   \title{A deep view on the Virgo cluster core}


   \author{S. Lieder
          \inst{1,2}
          \and
          T. Lisker\inst{1}
	  \and M.Hilker\inst{3}
	  \and I.Misgeld\inst{4,3}
	  \and P.Durrell\inst{5}
          }

   \institute{Astronomisches Rechen-Institut, Zentrum f\"ur Astronomie der
  Universit\"at Heidelberg, M\"onchhofstra\ss e 12-14, 69120
  Heidelberg, Germany, \email{slieder@ari.uni-heidelberg.de}
	      \and
	      European Southern Observatory, Av. Alonso de C\'ordova 3107, Vitacura, Santiago, Chile
	      \and
	      European Southern Observatory, Karl-Schwarzschild-Strasse 2, 85748 Garching bei M\"unchen, Germany
	      \and
	      Universit\"ats-Sternwarte M\"unchen, Scheinerstr. 1, D-81679 M\"unchen, Germany
	      \and
	      Department of Physics \& Astronomy, Youngstown State University, Youngstown, OH 44555, USA
             }

   \date{Received 2 May 2011; accepted 15 November 2011}

 
  \abstract
   {Studies of dwarf spheroidal (dSph) galaxies with statistically
     significant sample sizes are still rare beyond the
     Local Group, since these low surface brightness objects can only
     be identified with deep imaging data. In galaxy clusters, where
     they constitute the dominant population in terms of number, they
     represent the faint end slope of the galaxy luminosity function and 
     provide important insight on the interplay between galaxy mass
     and environment. In this study we investigate the optical photometric properties of early-type
     galaxies (dwarf ellipticals (dEs) and dSphs) in the Virgo
     cluster core region, by analysing their location on the colour
     magnitude relation (CMR) and the structural scaling relations
     down to faint magnitudes, and by constructing the luminosity
     function to compare it with theoretical expectations. Our work is based on deep CFHT V- and I-band data covering several square degrees of the Virgo cluster core that were obtained in
     1999 using the CFH12K instrument.\\
     We visually select potential cluster members based on morphology
     and angular size, excluding spiral galaxies. A photometric
     analysis has been carried out for 295 galaxies, using surface
     brightness profile shape and colour as further criteria to identify
     probable background contaminants. 216
     galaxies are considered to be certain or probable Virgo cluster members. Our study reveals 77 galaxies not catalogued in the VCC (with 13 of them
     already found in previous studies) that are very
     likely Virgo cluster members because they follow the Virgo CMR
     and exhibit low S\'ersic indices. Those galaxies reach
     $M_V=-8.7$ mag.\\
     The CMR shows a clear change in
     slope from dEs to dSphs, while the scatter of the CMR in the dSph regime does
     not increase significantly. Our sample might, however, be
     somewhat biased towards redder colours. The scaling relations
     given by the dEs appear to be continued by the dSphs indicating a similar origin. The observed change in the CMR slope may mark the point at which gas loss prevented significant metal enrichment.
 The almost constant scatter around the CMR possibly indicates a short formation period, resulting in similar stellar populations.\\
     The luminosity function shows a Schechter function's faint end slope of
     $\alpha=-1.50\pm0.17$, implying a lack of galaxies related to the
     expected number of low-mass dark matter haloes from theoretical
     models. Our findings could be explained by suppressed star
     formation in low-mass dark matter halos or by tidal disruption of dwarfs
     in the dense core region of the cluster.} 


   \maketitle
%

\section{Introduction}
The galaxy population of the Virgo cluster is well studied down to a B
magnitude of 19 mag by the photographic survey of \citet{binggeli}
which resulted in the Virgo cluster catalog (VCC). In addition,
\citet{impey}, \citet{tully}, \citet{trentham} and \citet{saba03} provided catalogs of
(very) low surface brightness objects in this region down to a B
magnitude of 21.5 mag. All these studies revealed a faint end slope of the LF in the
range
$-1.1\lesssim\alpha\lesssim\-1.6$. This also holds for different
nearby galaxy clusters (Fornax: \citet{ferguson, mies07},
Centaurus: \citet{misg09}, Hydra I: \citet{misgeld08}, Perseus:
\citet{penn07}, Coma: \citet{beijersbergen}, Abell 2199:
\citet{rine08}). However, simulations of Milky Way-sized halos
\citep{klypin,moore99} based on the currently favoured cosmological
cold dark matter ($\Lambda$CDM) model, in which galaxies form
hierarchically \citep{white}, show a CDM mass function's faint end slope
of $\alpha\approx-1.8$. As deduced by \citet{tully}, the LF faint end slope has to be steeper than that of the mass function, i.e. $\alpha<-1.8$. Hence, a large number of DM
halos are not observed. On the other hand, \citet{moore99} derive a
Virgo cluster subhalo mass function by inverting the LF data of
\citet{binggeli} using the Tully-Fisher relation \citep{tf77}. That
mass function agrees both in shape and amplitude with the subhalo mass
function of their CDM simulations of a Virgo-sized halo. Clearly, reproducing
the observed LF is a fundamental test for any viable theory of galaxy
formation.\\

To investigate, from the observational side, the formation history of
dEs and dSphs and the impact of mass and environment on their evolution, 
scaling relations of structure (surface brightness, luminosity, and
size) and colour are fundamental tools. 
For a sample of dEs and Es taken from various environments, \citet{Kormendy} reports almost perpendicular sequences of dEs and Es in the diagram of central surface brightness versus luminosity. This behaviour was often interpreted as evidence for different formation processes of dEs and Es. On the other hand, \citet{ferguson} find in their Fornax cluster data that even the brightest galaxies follow the same continuous linear relation as the dwarf galaxies. \citet{graham} show in their analysis that there is no dichotomy between Es and dEs. The different $\mu_0-M$ scaling relation of brightest galaxies is due to core evolution and therefore not related to formation mechanism. This is confirmed by the study of \citet{ferrarese} who see the perpendicular relation for only the cored galaxies in the Virgo cluster.\\
The colour-magnitude relation
(CMR) of galaxies is commonly used to interpret the overall stellar
population characteristics of galaxies and deduce their
formation. Many authors report linear CMRs of early-type galaxies in clusters
\citep[e.g.][]{mies07, lisker08,misgeld08, smca08, misg09}, with gradually
redder colours at increasing luminosity. This is commonly interpreted as a relation between mass and metallicity, in the sense that more massive
galaxies can self-enrich their gas, while dwarf galaxies lose it more
easily. This by itself would not imply a different
formation process of early-type giant and dwarf galaxies.
Recently, however, \citet{ferrarese} reported a curved CMR of
100 Virgo early-type galaxies investigated with the HST ACS Virgo cluster
survey \citep[ACS VCS,][]{cote04}. From an even larger sample based on
homogeneous SDSS photometry, \citet{janz09} reported an S-shaped
Virgo CMR, in which dEs and Es seem offset from each other and
connected through a transition region. Yet by comparing the
observations to model predictions, \citeauthor{janz09} showed that
distinct formation processes between dEs and Es are not necessarily
required to explain the nonlinear CMR shape. It is thus still a matter of
debate whether dwarf galaxies are merely the faint extension of giant
galaxies or whether their origin is different.\\

Our main goal is to identify and analyse dwarf galaxies in the
vicinity of the three giant ellipticals M87 and M86/M84 in the Virgo cluster core, in a similar manner
 as it was done by our group for the Fornax cluster \citep{hilk99, hilk03, mies07} and the Hydra I \citep{misgeld08} and Centaurus clusters \citep{misg09}. We aim to investigate the Virgo cluster luminosity
function (LF) down to a V band magnitude of 22, which is 
approximately two magnitudes lower than the mentioned Virgo cluster
studies. \citet{grebel01} defines the luminosity
$M_V\approx-17$ mag as dividing parameter between giant and dwarf
early-type galaxies. The early-type dwarf galaxies are commonly
subdivided into two classes, the dwarf ellipticals (dEs) and the
fainter dwarf spheroidals (dSphs). The latter are distiguished from
the former by a dividing luminosity of $M_V\approx-14$ mag (see
\citealt{grebel01}). The diameter of the Virgo cluster is approximately 3 Mpc and its crossing
time is $\sim0.1 H^{-1}_0$ \citep{tull96}, so that its galaxies have
had time to interact with each other. But the Virgo cluster is
considered as dynamically young cluster, whereas the dynamically old central region is surrounded by a not virialized but infalling region \citep{bing93}. This is underlined by the irregular structure of at least
three subclusters (centered on M87, M86 and M49), suggesting the Virgo cluster might be a complex
unrelaxed system \citep{saba03}.


\section{Data \& visual inspection}
   \begin{figure}
   \centering
   \includegraphics[width=0.5\textwidth]{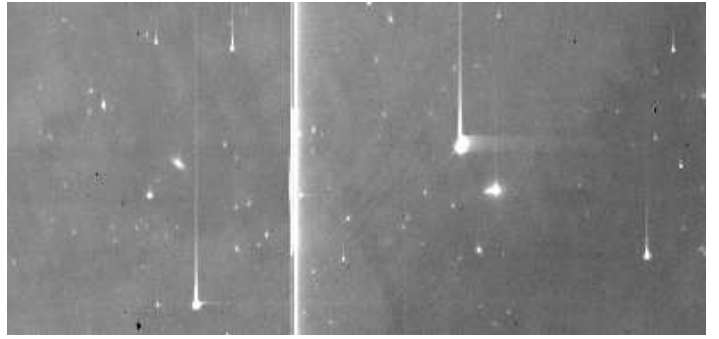}
      \caption{Excerpt of the defective chip to illustrate read-out defects.}
         \label{fig:defect}
   \end{figure}     

   From May 1st to May 5th 1999, V- and I-band CFHT wide field imaging
   data of the central region of the Virgo cluster around M86 and M87
   were acquired using the CFH12K camera at the Canada-France-Hawaii Telescope (CFHT). The CFH12K camera was a mosaic of 12 CCDs, with  a pixel scale of 0.206 arcsec pix$^{-1}$ and a total areal coverage of  $28'\times42'$ per field. However, only 11 of the 12 CCDs were functioning properly -- the CCD on the bottom right corner of the layout showed an extended hot region
(mostly along the y-axis) and significant read-out defects in the
sense that all source detections exhibit a long additional trail in
y-direction (see Fig. \ref{fig:defect}). Thus, data of this chip was
excluded from the analysis.\\
  \begin{figure}
   \centering
   \includegraphics[width=0.5\textwidth]{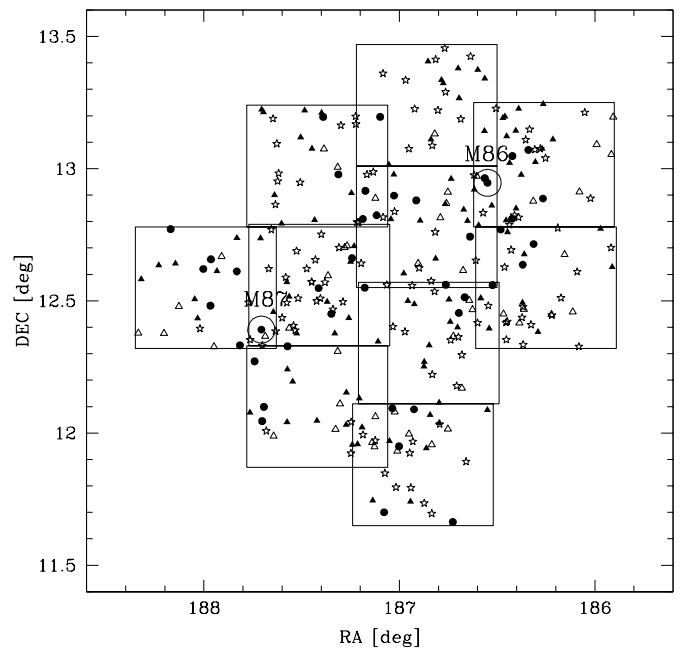}
      \caption{Coordinate map of the 296 photometrically investigated objects. Filled circles: redshift confirmed cluster members, open triangles: redshift confirmed background galaxies, filled triangles: remaining galaxies catalogued in VCC, stars: no information. The well known giant ellipticals M86/M87 are highlighted, rectangles: fields covered by CFH12K instrument}
         \label{fig:coord}
   \end{figure}   
A total of 10 fields were obtained, covering a region of $\approx 1.5 \times 2.5$ square degrees, as shown in Figure \ref{fig:coord}. A small overlap of the fields ensured a
   complete coverage of the observed Virgo core region. Every field
   was exposed at least three times with an exposure time of 600 seconds
   each. The field, containing M84/M86 was exposed four times, the one containing M87 even five times, resulting in an overall exposure time of 50 minutes. In case of saturated galaxies in the long exposures
   additional 60s exposures were taken. During the four observing
   nights the median seeing was 0.8 arcsec FWHM.

\subsection{Data reduction}
   The THELI image reduction pipeline \citep{erben}
   was used for the preprocessing of the data. All of the CFH12K data
   were corrected for the bias levels and the dark current, and were
   flat-fielded using both twilight flats and superflats made from all
   33 science images. For the superflat creation we excluded images of
   CCDs which covered giant galaxies because of the small applied
   dithering of our data. By defringing the superflats, the brickwall pattern \citep{cfht} present in the data was almost
   completely removed. The astrometric calibration from the THELI
   reductions is based on cross-correlation with the SDSS catalog of
   point sources, which also corrected for geometric distortions in
   the outer most parts of CFH12K fields. After the THELI photometry
   step, which is based on SExtractor \citep{bertin}, background subtraction was
   carried out using THELI, except in those cases where extended
   bright objects were present. In those cases, background subtraction
   was carried out manually. 
After the THELI processing, instrumental magnitudes were computed from observations of standard stars taken in all four nights of the observing run, and the photometry calibrated on the Cousins $V$ and $I$ magnitude system of \citet{landolt}.\\
   The average noise per pixel for the 30 minutes exposed fields corresponds to surface brightnesses
   of $\mu_V=26.5$ mag/arcsec$^2$, $\mu_I=25.2$ mag/arcsec$^2$
   respectively. Hence, the V-band is in general deeper than our
   I-band observations.

\subsection{Visual inspection}\label{sec:inspection}
   All images were carefully inspected visually to detect possible Virgo cluster members. The images in both bands were inspected independently from each other. A first pre-selection was based on the following selection criteria.\\
   \begin{enumerate}
   \item The depth of the observed data revealed many background
     objects. While their physical size and luminosity would typically
     be larger than that of low-mass Virgo galaxies, their apparent
     size and magnitude can be similar, so that background objects can
     be mistaken for a Virgo cluster galaxy. Thus, we limited the
     visual size (at a surface brightness of $\mu_V\approx 26$
     mag/arcsec$^2$) of objects that are taken into account to a
     radius $\approx 10$ arcseconds (see Fig. \ref{fig:crit2}). This
     is comparable to the size of the smallest galaxies listed in
     \citeauthor{binggeli}'s (\citeyear{binggeli}) Virgo cluster catalog (VCC) -- denoted by
     $D_{25}$ -- the diameter at a surface brightness of
     $\mu\approx 25$ mag/arcsec$^2$ in B.
   \item  Spiral galaxies were not considered in this work. In the literature very few dwarf spirals are reported (e.g. \citealp{schombert}, \citealp{graham}). Therefore we excluded all faint small galaxies with obvious spiral structure, because they are regarded as background galaxies (see Fig. \ref{fig:crit3}). Since we are using the \verb+IRAF ellipse+ task for photometry, also giant spirals are excluded because of the bad modelling by ellipses. The photmetric errors would be to large. 
   \item Irregular galaxies were considered as long as their shape has been tolerably elliptical to be modelled with \verb*#ellipse# (see Fig. \ref{fig:crit4}). Four dIrrs were considered.
   \end{enumerate}
   Particularly because of the second criterion, we mainly obtain early-type galaxies. In particular, at faint magnitudes ($M_V>-13$ mag) we deal only with early-type galaxies. The criteria lead to a sample of 371 selected objects.

\section{Photometry}

  \subsection{Photometric analysis}
   All objects were photometrically analysed using the \verb*#ellipse# task \citep{ellipse} which is included in the \verb*#STSDAS# package of \verb*#IRAF#. All \verb*#ellipse# fits were performed with fixed parameters for center coordinates, position angle and ellipticity. In some cases like M86 and M87 better fitting results were obtained when variable ellipticity was applied. Obvious foreground and background objects were masked using the interactive mode of \verb*#ellipse#. \\
   The
   \verb*#ellipse# output tables have been used to determine all astronomical quantities which are presented in this study. The extent of a galaxy has been defined to be the isophote at which the intensity falls below 10\% of the RMS variation of intensity \emph{along} the isophote\footnote{The RMS value is determined by ellipse, subdividing the isophote into sectors where appropriate \citep{ellipse}.. The so determined radii are of the order of 2 to 3 effective radii $R_e$ for dSphs and 3 to 4 $R_e$ for the dEs.} 
The flux enclosed by that ellipse is used as total flux $f$. Using that flux the apparent magnitude of an object is calculated. Finally, the apparent magnitude of an object was corrected for galactic extinction, applying the results of \citet{schlegel}.\\
   In Tab. \ref{table:calibration} quantities determined for the photometric calibration are listed. At high luminosities the photometry is limited by the uncertainty of the zeropoint and the atmospheric extiction coefficient $\kappa$. The photometrical uncertainties of the dSphs are additionally affected by the sky noise which is then of the same order of magnitude or higher as compared to zeropoint and $\kappa$.

\begin{table*}
\caption{Photometric calibration quantities.}             
\label{table:calibration}      
\centering                          
\begin{tabular}{c c c c c}        
\hline\hline                 
Filter & ZP [mag] & $\kappa$ [mag] & X & A [mag]\\    
\hline                        
V & $26.245\pm0.064$ & $-0.086\pm0.048$ & $(1.019\dots1.147)\pm(0.00\dots0.06)$ & $0.064\dots0.198$ \\
I & $26.136\pm0.038$ & $-0.039\pm0.029$ & $(1.030\dots1.203)\pm(0.00\dots0.09)$ & $0.038\dots0.114$ \\
\hline                                   
\end{tabular}\\
\tablefoot{ZP: Zeropoint, $\kappa$: atmospheric extiction coefficient, X: mean airmass of exposures contributing to a coadded image, A: galactic extinction by \citet{schlegel}} \\
\end{table*}

\subsection{Surface brightness profiles}
   \begin{figure}
   \centering
   \includegraphics[width=0.5\textwidth]{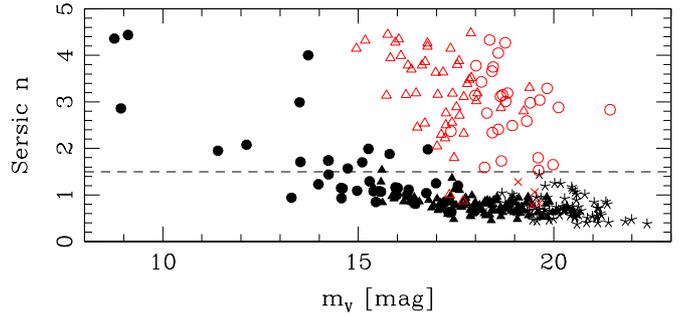}
      \caption{S\'{e}rsic indices of all investigated objects versus apparent V-Band magnitude. Black circles: spectroscopically confirmed Virgo members, black triangles: VCC members without redshift information, red triangles: spectroscopically confirmed background objects, red circles: as background considered objects due to their large S\'{e}rsic index, black asterisks: as Virgo member considered objects due to their small S\'{e}rsic index, red crosses: objects with small S\'{e}rsic index but extreme red colour (see Sect. \ref{sec:cmd}), dashed line: Virgo cluster membership dividing S\'{e}rsic index of 1.5 (for magnitudes fainter than $m_V\approx18$ mag)}
         \label{fig:sersic}
   \end{figure}   

The half-light radius $r_{50}$ is determined as the radius enclosing
50\% of the total flux. The effective surface brightness $\mu_e$ is
defined as the average surface brightness within $r_{50}$. We note
that, due to our above definition of the total flux,
$r_{50}$ differs slightly from the true effective radius, which would enclose
50\% of the total flux obtained by integrating the light profile to
infinity.

   Spectroscopic data are not available for many of the (largely) faint objects in our sample. As a result, we will need to use the surface brightness profiles of these objects in order to distinguish true Virgo cluster members from background galaxies (see Sect. \ref{sec:sample}). We performed single profile S\'{e}rsic fits \citep{sersic} to all objects, that is
\begin{equation}
I(r)=I_e\exp{\left\{-b_n\left[\left(\frac{r}{R_e}\right)^{1/n}-1\right]\right\}}.
\end{equation}
$I_e$ the intensity of the isophote at the effective $R_e$. The constant $b_n$ is defined in terms of the parameter $n$ which describes the shape of the light profile. As shown by \citet{caon93}, a convenient approximation
relating $b_n$ to the shape parameter n is $b_n=1.9992n-0.3271$ for
$1\lesssim n\lesssim10$, which we applied in our calculations. For the
S\'{e}rsic fit the inner 3 arcsec were excluded from the fit which is
twice the worst seeing with a FWHM of 1.5 arcsec. For nucleated
galaxies only the main body of the galaxy was fitted. The results of
this analysis, displayed in Fig. \ref{fig:sersic}, show that at
intermediate magnitudes ($15\lesssim m_V\lesssim18$ mag) the
confirmed background  galaxies (red open triangles) significantly
differ in their S\'{e}rsic index from confirmed Virgo cluster members
(black filled circles) in the same magnitude range. High S\'{e}rsic
indices are typical for Virgo members with $m_V\lesssim15$ mag. The
red open circles in this plot show a S\'{e}rsic index distribution
similar to that of the confirmed backgound galaxies, but exhibiting
lower apparent magnitudes. Since those objects have large values of
$n$ (typical for giant ellipticals) we therefore conclude, that these
objects are distant background elliptical galaxies. In the plot the VCC members without
redshift information (black filled triangles) continue the trend given
by confirmed members, resulting in low $n$ (typical for dwarf
galaxies). There is a further group of galaxies with low $n$ (black
asterisks) continuing the trend of the VCC galaxies at lowest
luminosities ($m_V\gtrsim19$). We conclude, that these diffuse objects
(characterized by their small $n$) also belong to the Virgo
cluster. Furthermore there is a noteworthy clean separation between
the two groups of unknown objects (red open circles and black
asterisks). Thus, we adopted a membership criterion of $n<1.5$ (for
$m_V\gtrsim15$ mag), denoted by the dashed line in Fig. \ref{fig:sersic}.


\section{Sample selection and subdivision}\label{sec:sample}

   \begin{figure*}
   \subfigure[very low surface brightness object which passed the
     selection
     criteria]{\includegraphics[width=0.23\textwidth]{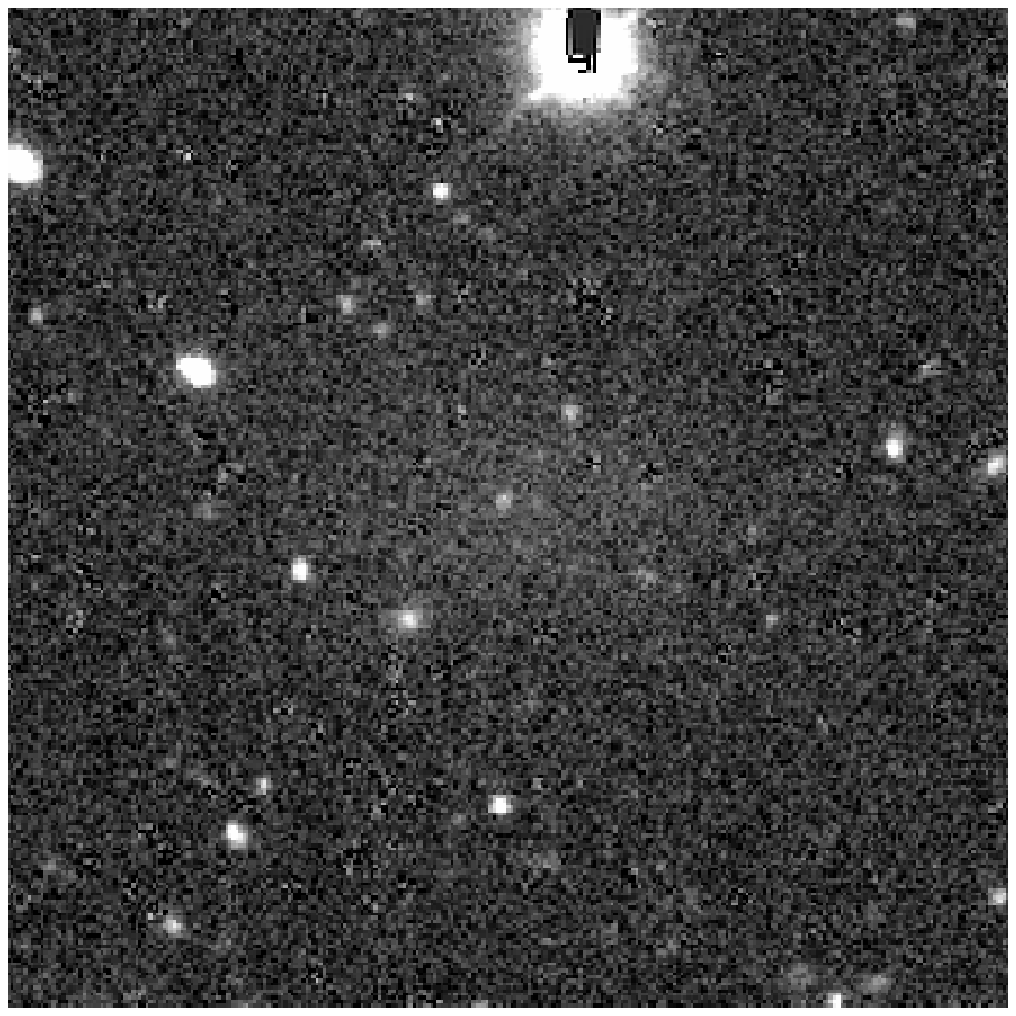}\label{fig:crit1}
     }\hfill
   \subfigure[rejected object due to size criterion (yellow circle
     denotes a 10"
     diameter)]{\includegraphics[width=0.23\textwidth]{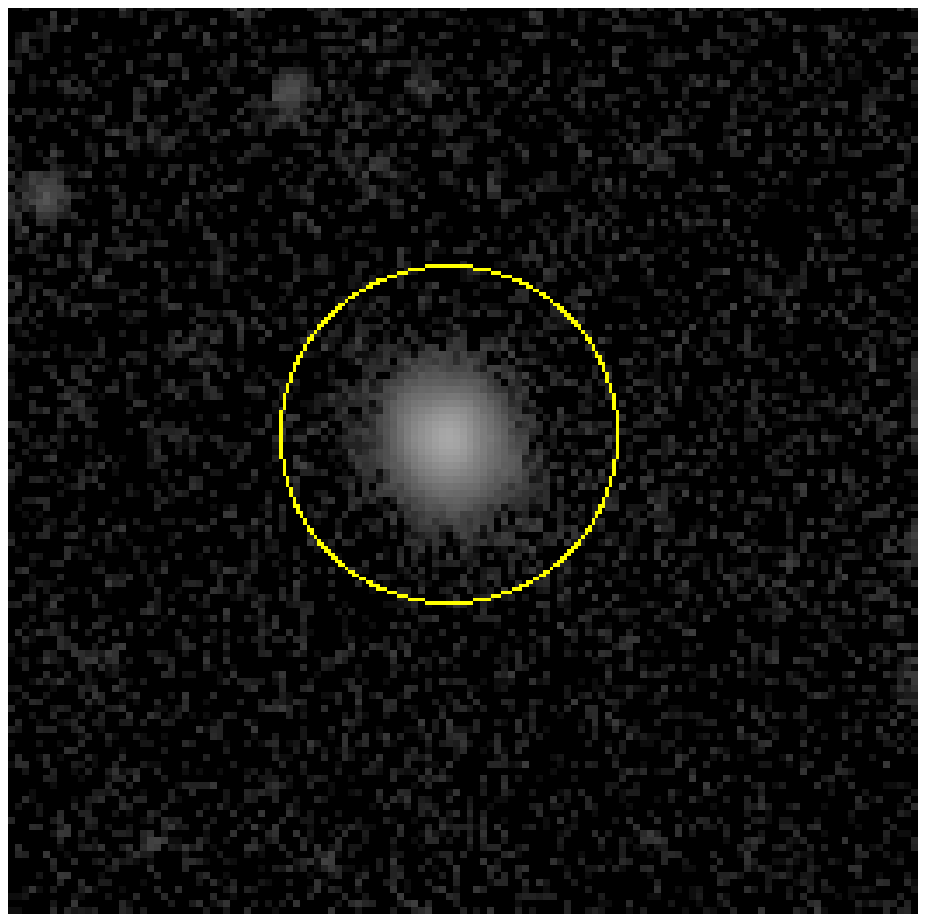}\label{fig:crit2}
    }\hfill
   \subfigure[rejected galaxy due to spiral criterion (yellow circle
     denotes a 10"
     diameter)]{\includegraphics[width=0.23\textwidth]{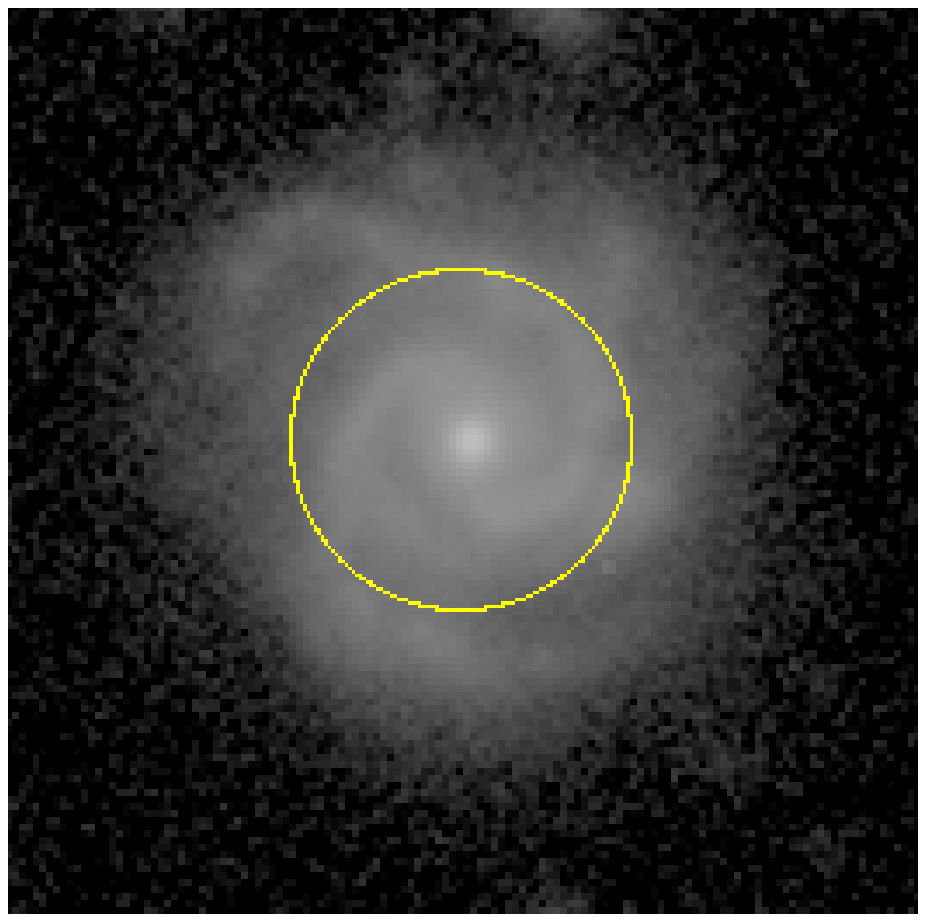}\label{fig:crit3}
    }\hfill
   \subfigure[irregular galaxy rejected due to its
     shape]{\includegraphics[width=0.23\textwidth]{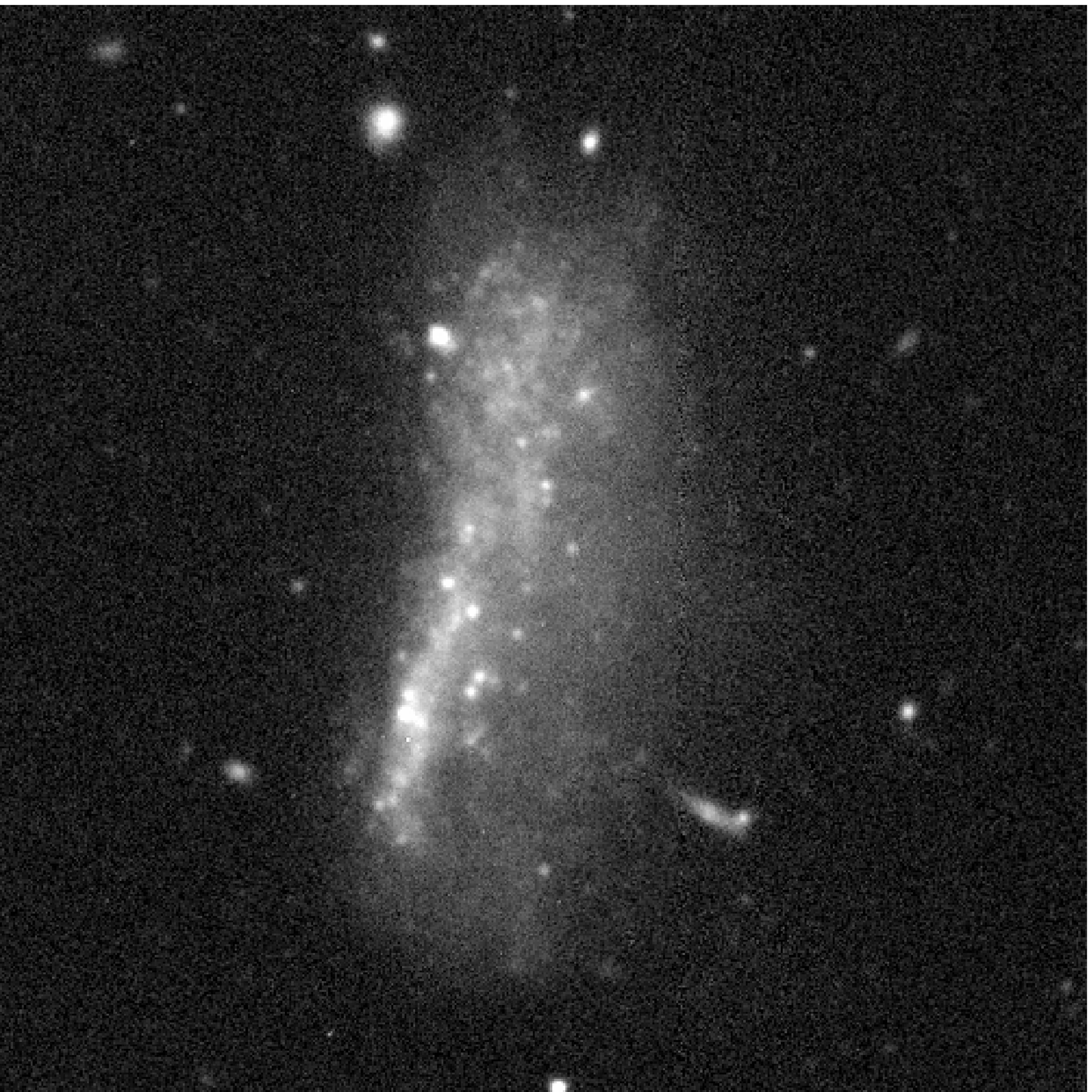}\label{fig:crit4}
    }\hfill
   \subfigure[object with heavy spike in central
     region]{\includegraphics[width=0.23\textwidth]{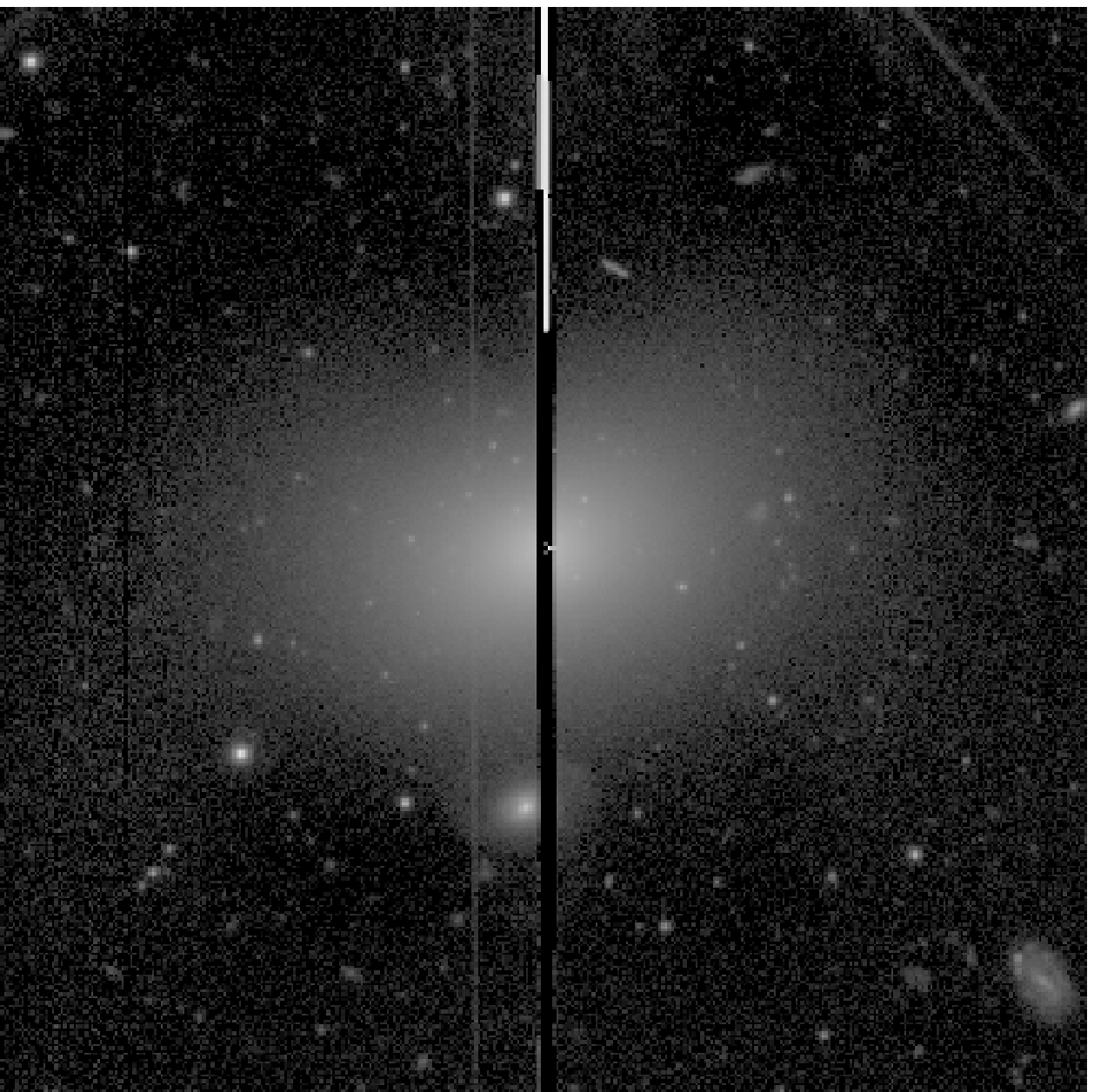}\label{fig:crit5}
    }\hfill
   \subfigure[object located in a foreground star's
     halo]{\includegraphics[width=0.23\textwidth]{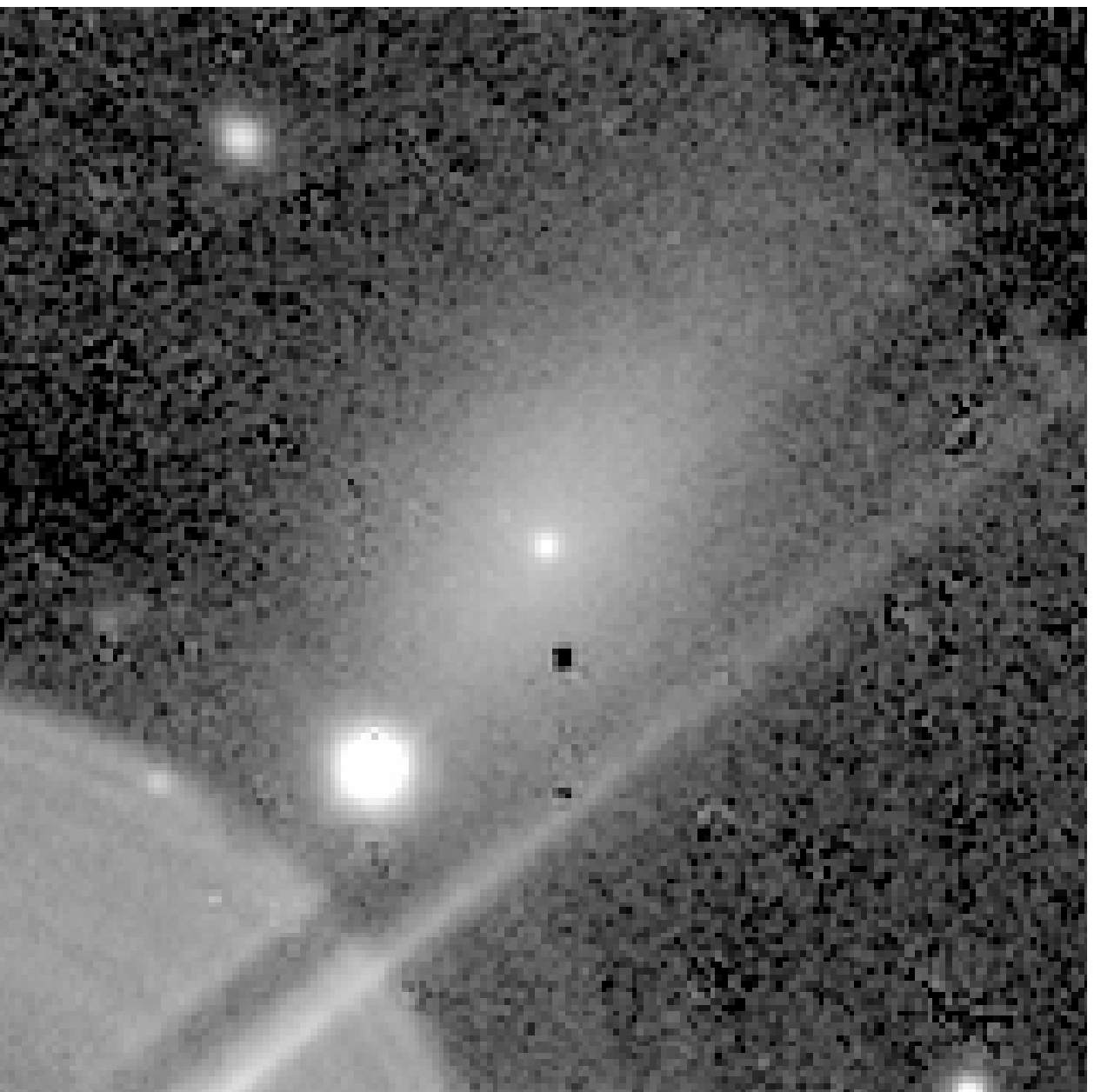}\label{fig:crit6}
    }\hfill
   \subfigure[object with merger
     signature]{\includegraphics[width=0.23\textwidth]{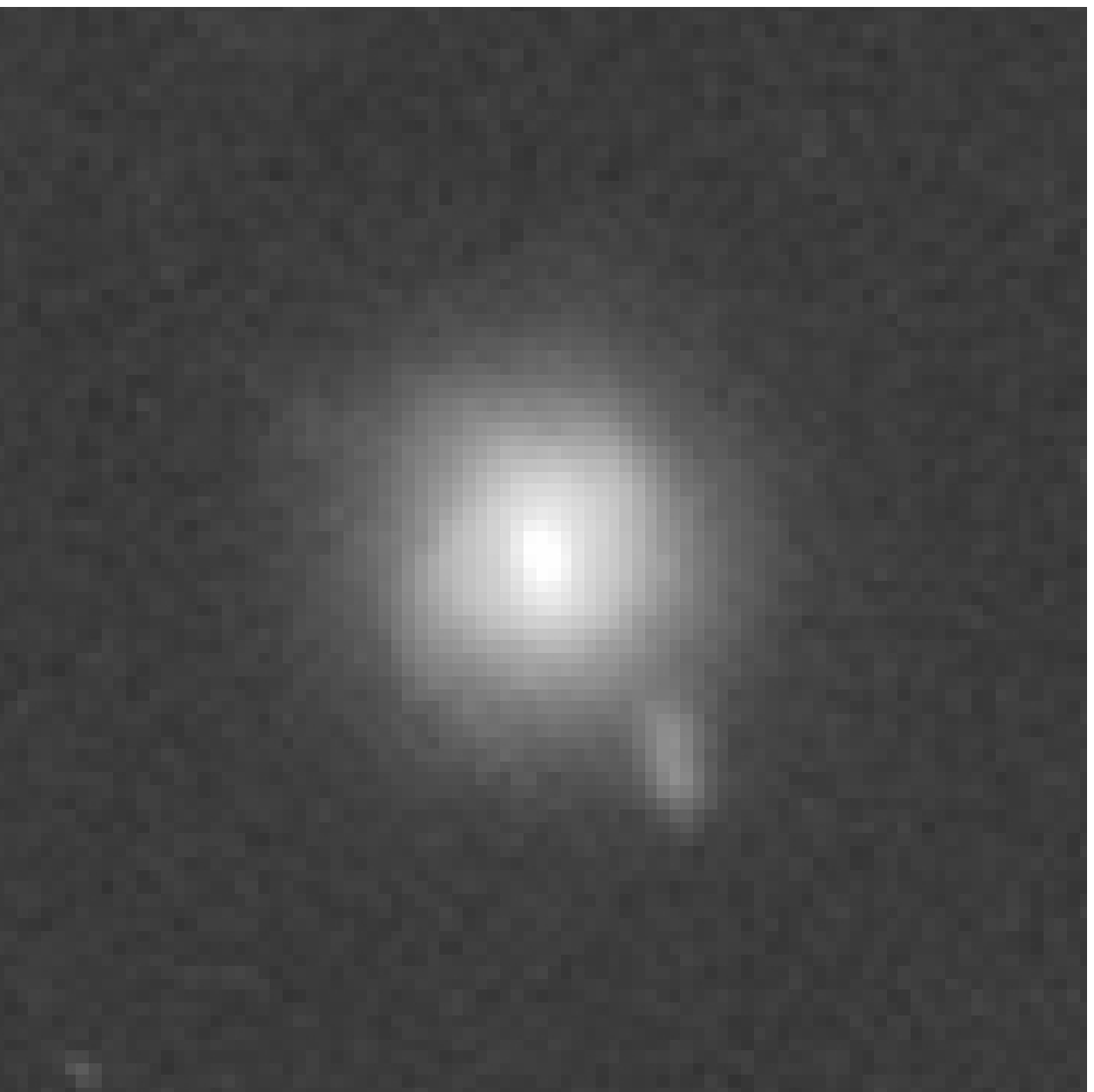}\label{fig:crit7}
    }\hfill
   \subfigure[object \ref{fig:crit7} after after best ellipse model
     subtraction]{\includegraphics[width=0.23\textwidth]{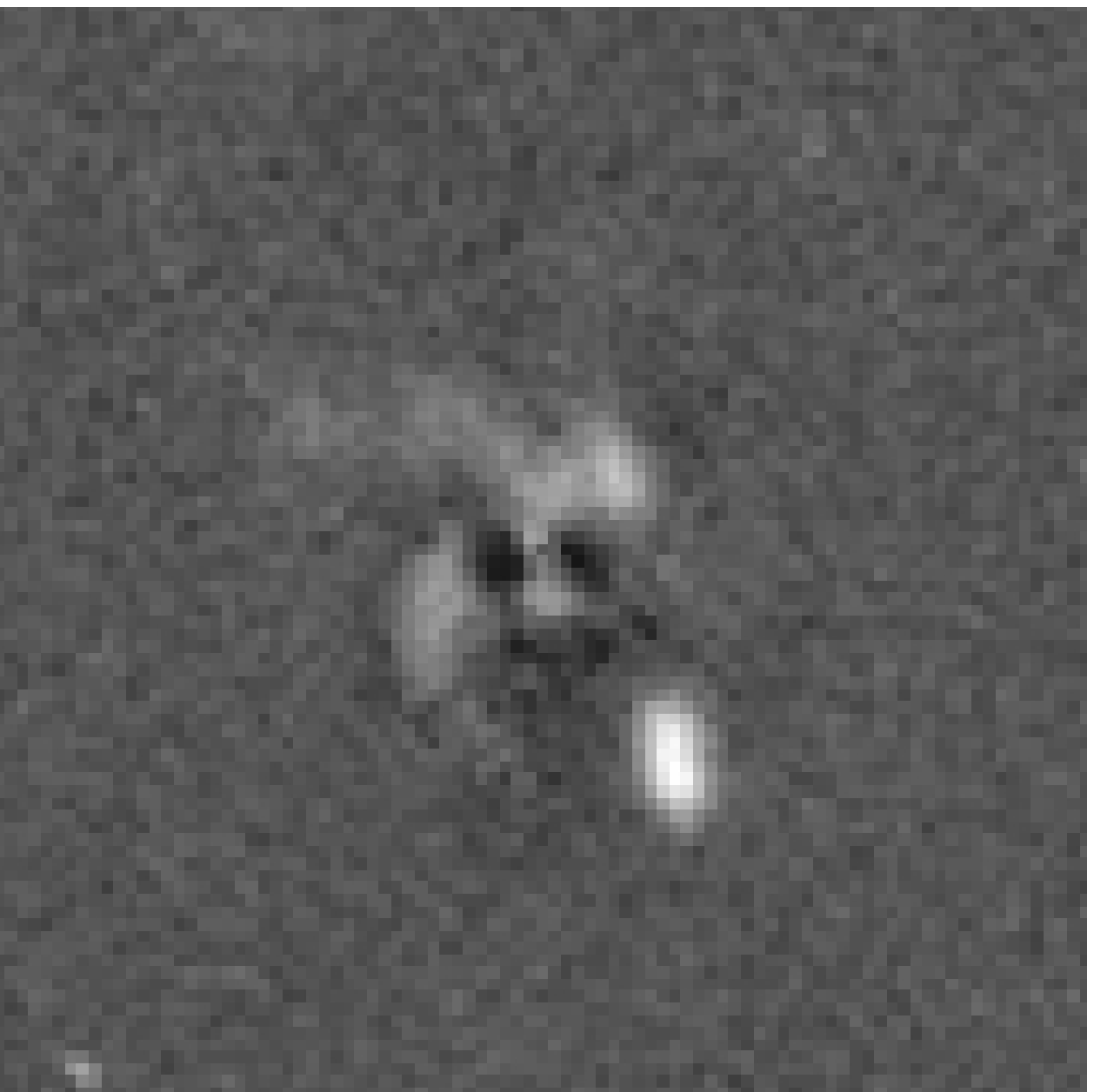}\label{fig:crit8}
    }\hfill
   \subfigure[object with spiral
     structure]{\includegraphics[width=0.23\textwidth]{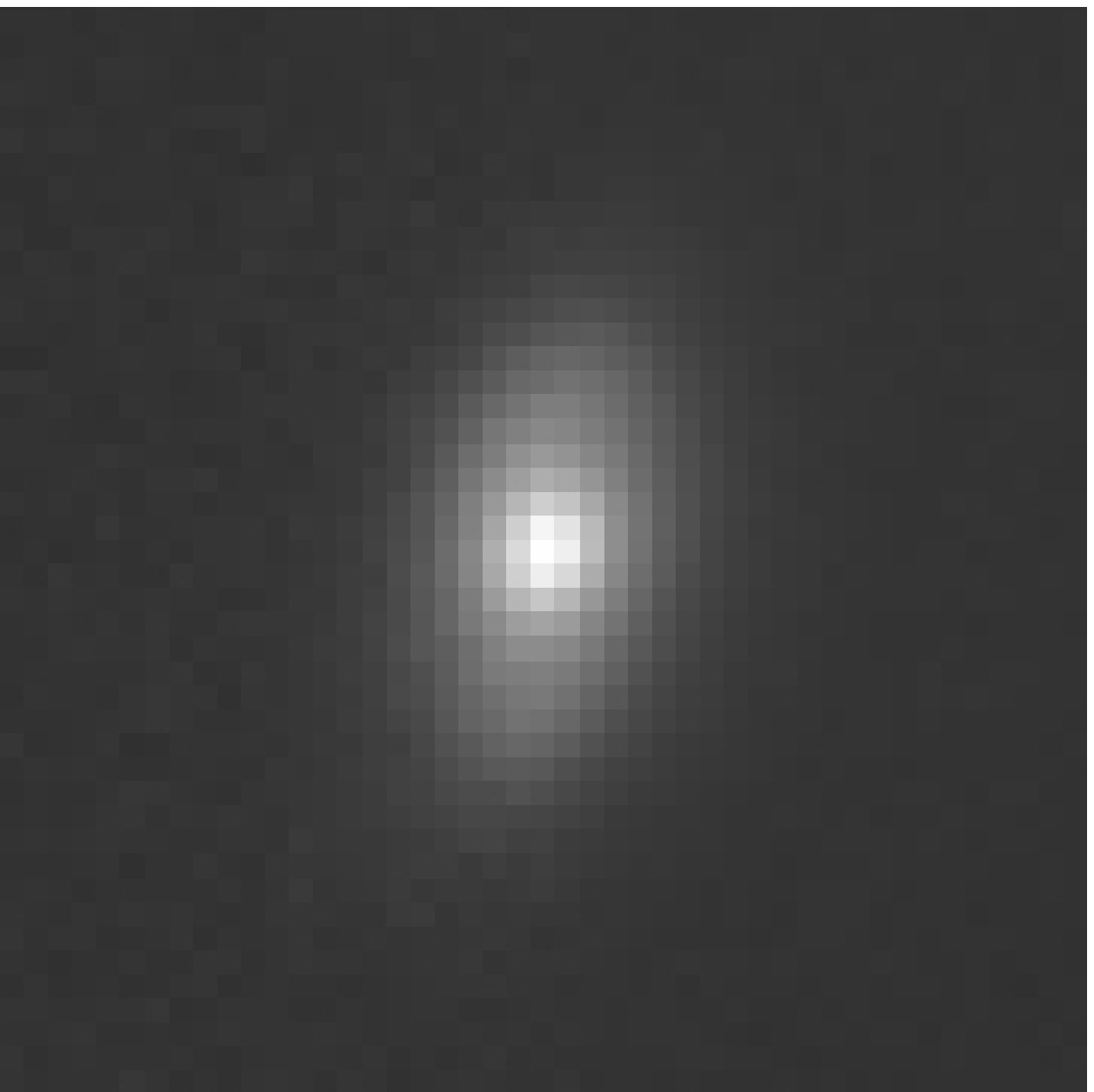}\label{fig:crit9}
    }\hfill
   \subfigure[object \ref{fig:crit9} after best ellipse model
     subtraction]{\includegraphics[width=0.23\textwidth]{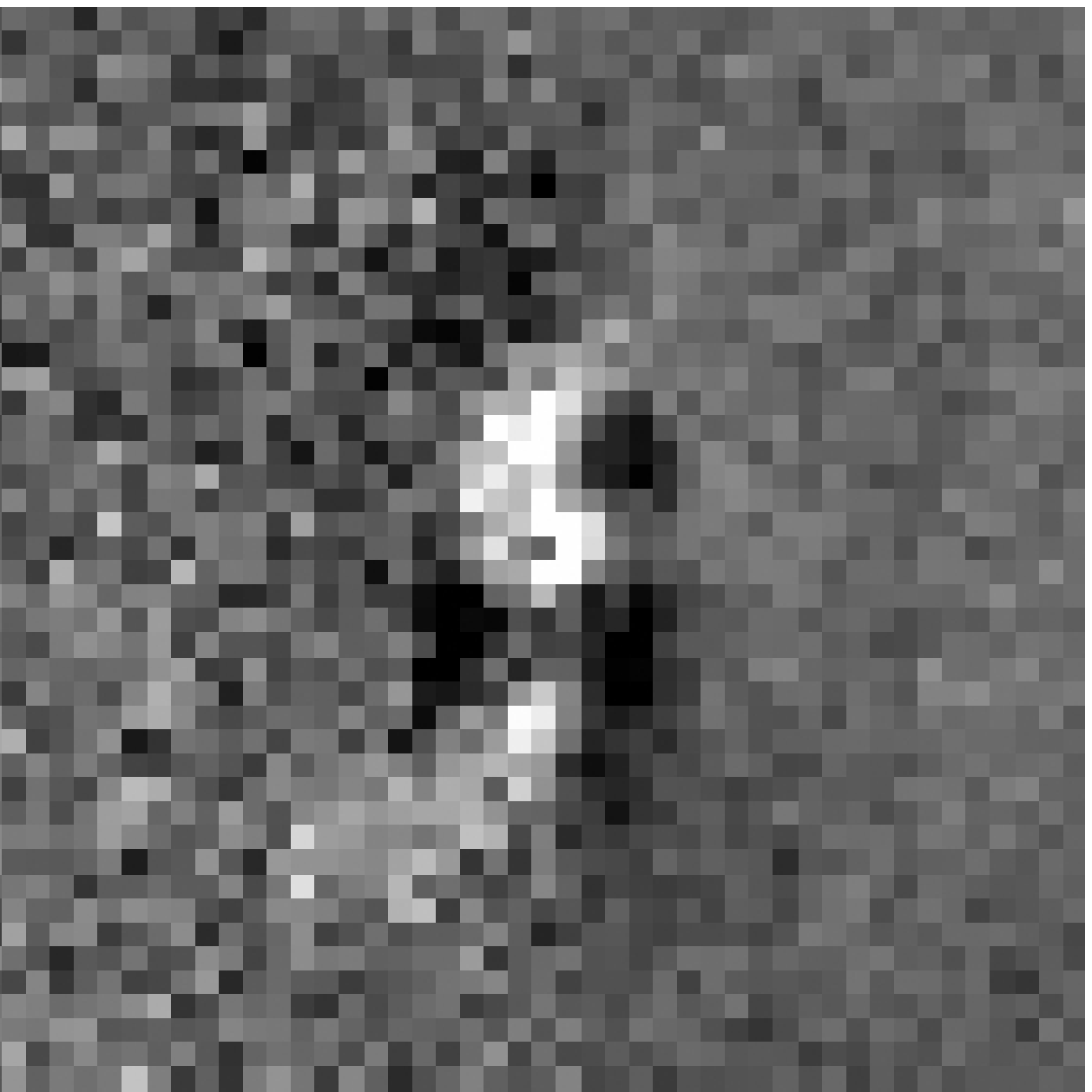}\label{fig:crit10}
    }\hfill
   \subfigure[object which is thought to be a spiral galaxy or an E/S0
     because of the apparent dust
     signature]{\includegraphics[width=0.23\textwidth]{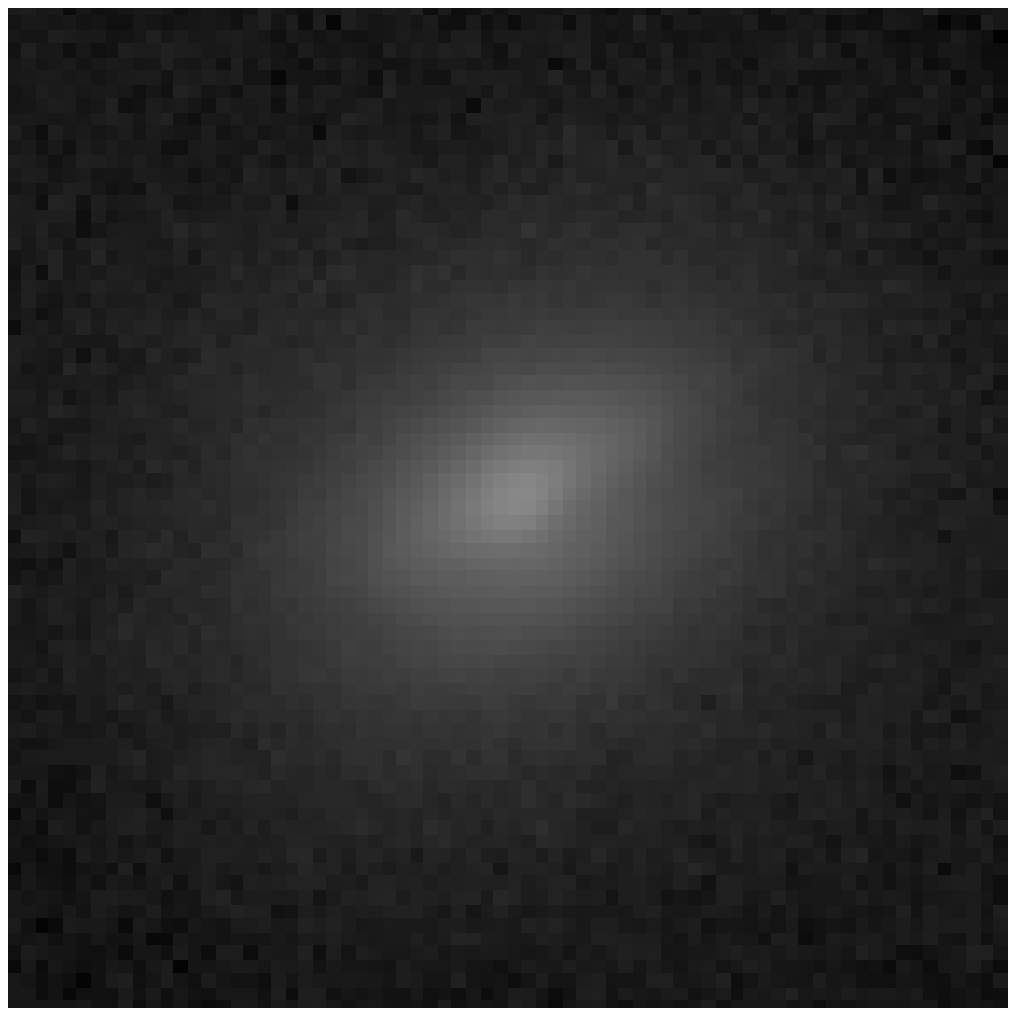}\label{fig:crit11}
    }\hfill
   \subfigure[rejected compact galaxy located on the Virgo red
     sequence (possible
     cE).]{\includegraphics[width=0.23\textwidth]{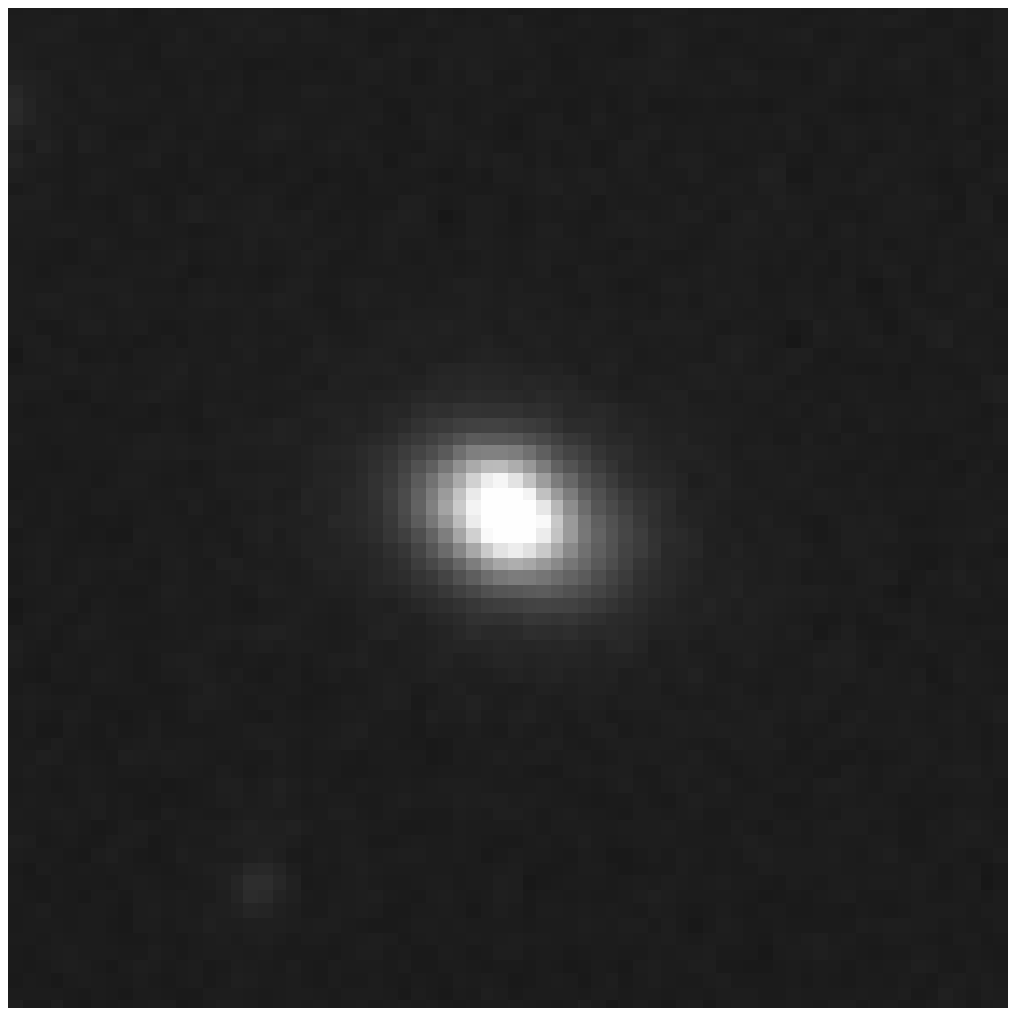}\label{fig:crit12}
    }\hfill
   \caption{Figures for clarification of selection criteria}
   \label{fig:criteria}
   \end{figure*}

During the photometric analysis using \verb*#ellipse# 75 objects (from the initial list of 371 objects) were rejected for any of a number of reasons described below (some of which are illustrated in Fig. \ref{fig:criteria}):
   \begin{enumerate}
   \item The photometry was affected by the presence of either a diffraction spike or bled columns from bright or saturated stars (see Fig. \ref{fig:crit5}, 2 objects rejected). Masking the spike led to inappropriate \verb*#ellipse# fits.  
   \item The object was located within the "halo" of a foreground star. An appropriate \verb*#ellipse# fitting was impossible due to the partial overlap of the galaxy with the star's halo. (see Fig. \ref{fig:crit6}, 5 objects excluded)
   \item The subtraction of the best \verb*#ellipse# model from the original image revealed a merger signature. In this case, the light origins from two objects and an \verb*#ellipse# fit is not appropriate for the analysis. (see Figs. \ref{fig:crit7} and \ref{fig:crit8}, 5 objects excluded)
   \item The subtraction of the best \verb*#ellipse# model from the original image of small galaxies revealed a slight spiral structure. Since spirals are considered to be giants (see Sect. \ref{sec:inspection}), these objects are treated as background galaxies. (see Figs. \ref{fig:crit9} and \ref{fig:crit10}, 27 objects excluded, 12 of which are spectroscopically confirmed background galaxies)
   \item Some small galaxies show dust signatures (typical for giant E or S0) but the residual image of the best \verb*#ellipse# model subtraction does not show significant structures. In this case we trust our eyes and classify the galaxies as background. (see Fig. \ref{fig:crit11}, 11 objects excluded)
   \item An object was only taken into account if it was visible in both bands, V and I. (5 objects excluded)
   \item Very faint, diffuse galaxies lead to unrobust fits. (3 objects excluded)
   \item There were too many bad pixels within the object or the object was only partly covered. (3 objects excluded)
   \item An appropriate local background subtraction was impossible because of small scale spatial brightness variations. A residual fringe pattern remained on one chip, and some objects were also close to a spiral galaxy or a tail of a merging galaxy, which both could not appropriately be subtracted. (5 objects excluded)
   \item Some initially visually selected objects were rejected because closer inspection revealed a too small size (diameter $<5$ arcsecs; 9 objects excluded)

   \end{enumerate}

A coordinate map of the entire remaining sample of the 295 investigated objects is displayed in Fig. \ref{fig:coord}. The NASA extragalactic database (NED) was searched in order to extract spectroscopic redshifts. The subsamples of different redshift information are denoted by different symbols in the figure. For Virgo cluster membership we adopted a heliocentric velocity range of $-900$ km/s $\leq v_r \leq 2700$ km/s, supported by Fig. 2 of \citet{mei}. The matching yielded 41 spectroscopically confirmed Virgo cluster members and 47 confirmed background galaxies. Four of the confirmed Virgo cluster members have not been catalogued by Binggeli's VCC. On the contrary, three of the background galaxies do have a VCC number\footnote{\citet{binggeli} also included background galaxies and classified them. Our three excluded galaxies were classified as member (2) and possible member (1).}. The remaining 207 objects did not have any spectroscopic information. We have assumed that the 98 objects that are listed in the VCC are bona fide Virgo members. 
After these criteria, there remains a total of 109 objects not
catalogued in the VCC, some of which are likely contaminating background galaxies. From the previous section, we consider 28 of these objects with S\'{e}rsic index $n>1.5$ as background galaxies. Finally we use the integrated $V-I$ colours (see Sect. \ref{sec:cmd}) to remove four objects which have extremely red colours. This leads to a sample of 77 faint diffuse galaxies without redshift information which follow the CMR of the Virgo cluster. The properties of these probable Virgo dwarf galaxies are listed in Table \ref{table:appendix}. 13 of them have previously been found in other studies (\citealp{trentham,tully,durrell,impey,durr97}).\\
   In the following, we combine the sample of the 41 spectroscopically confirmed Virgo members, the 98 VCC members without redshift information and the 77 probable member galaxies. This combined Virgo cluster sample contains 216 galaxies.

\section{Results}
  Because the data is of higher quality in V-band we present all
  results related to the according V-band quantity, resulting in
  smaller errors. Furthermore it is mentioned here, that potential
  systematic errors due to local background subtraction are not
  included in our estimates.
To convert apparent to absolute magnitudes, we adopt the Virgo
cluster distance determined by the ACS VCS ($m-M=31.09$
mag, $d=16.5$ Mpc, \citealt{mei}) and use this value for all our galaxies
throughout the paper.

  \subsection{Colour magnitude diagram}\label{sec:cmd}
   \begin{figure}
   \centering
   \includegraphics[width=0.5\textwidth]{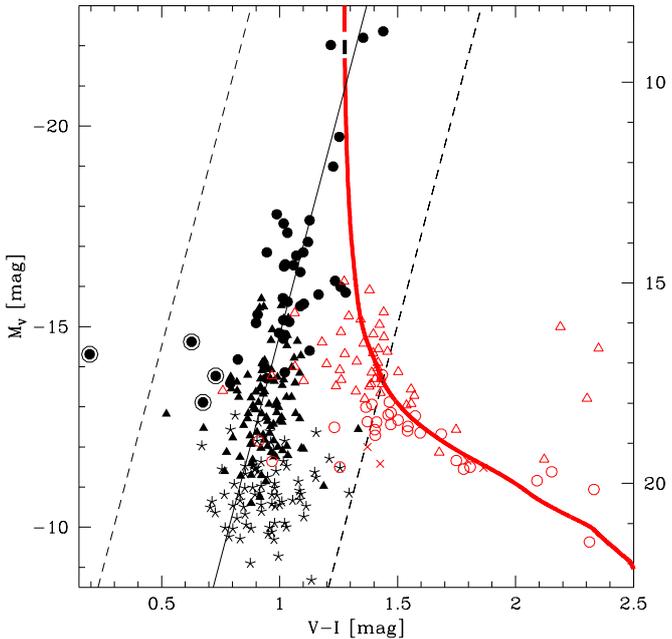}
      \caption{CMD of all investigated objects. Filled circles: spectroscopically confirmed Virgo cluster members, filled circles with surrounding open circle: irregular Virgo galaxies (confirmed), open triangles: spectroscopically confirmed background objects, filled triangles: remaining objects with VCC index, open circles: remaining objects with S\'{e}rsic index $n>1.5$, asterisks: assumed Virgo cluster members, crosses: excluded from Virgo cluster sample by position in CMD. Thin black line: fit to confirmed cluster members; dashed lines: $5\sigma$ confidence interval; thick line: redshift evolution of an E-type galaxy modelled by GALEV (dark grey intercept: Virgo redshift, red: redshifted up to $z=0.65$)}
         \label{fig:cmd}
   \end{figure}   
   Since spectroscopic data were not available for most of the faint objects, we also investigated obvious background objects with a comparable apparent magnitude as the Virgo member galaxies to determine their location in the colour magnitude diagram (CMD). The resulting CMD for all investigated objects is shown in Fig. \ref{fig:cmd}. The diagram shows a clear distinction between background (red open triangles and circles) and Virgo objects (black filled circles and triangles). We use that distinct occupation in the CMD to reject possible background objects from the sample of faint diffuse objects (asterisks). For the determination of the early-type galaxies' CMR we exclude the four irregular galaxies. The CMR of the confirmed Virgo members (black filled circles) is denoted by the black solid line. Confirmed members, observed in our field of view, cover a magnitude range of $-22.3\lesssim M_V\lesssim -13.1$ mag. The integrated colour $V-I$ was determined at the half-light aperture of each galaxy. This early-type CMR (red sequence) is given by the linear fit
   \begin{equation}
	(V-I)_{\mathrm{50}}=(-0.045\pm0.007)\cdot M_V+(0.337\pm0.121)
   \end{equation}
   with a RMS of $0.098$. The obtained CMR is comparable to other studies of the Virgo cluster (see e.g. \citealp{lisker08}) and other nearby clusters (see e.g. \citealp{misgeld08,misg09}).\\
   The dashed lines represent the $5\sigma$ confidence interval. Four faint diffuse objects (flagged by crosses, see also Fig. \ref{fig:sersic}) are extreme outliers ($>5\sigma$) with respect to the CMR of the Virgo cluster galaxies and follow the trend given by the objects with large S\'{e}rsic index (red open circles), which are considered to be background objects. Hence, we define them to be background objects.\\
   \begin{figure}
   \centering
   \includegraphics[width=0.5\textwidth]{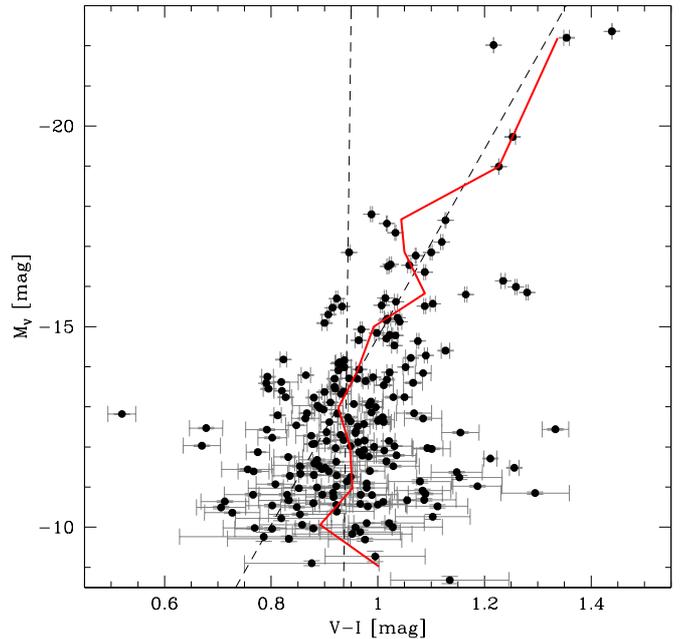}
      \caption{CMD of only the early-type Virgo sample is shown. The solid red line is constructed by the average measured in successive magnitude bins (1 mag per bin), the dashed black lines denote the CMR for the $M_V>-14$ mag subsample, $M_V<-14$ mag respectively. The errorbars shown here do not include the systematic uncertainty of the photometric zeropoint, as this would only cause an equal shift of all data points.}
         \label{fig:cmd2}
   \end{figure}   
   The CMD of our Virgo cluster sample obviously shows  a change in the slope of the CMR at $M_V\approx-14$ mag, as can be seen in Fig. \ref{fig:cmd2}. The red solid line indicates the average trend as found in successive magnitude bins with a width of $1$ mag and steps of $0.5$ mag, clipped one time at $3\sigma$ as performed in \citet{janz09}. We performed linear fits to both parts of the sample with $M_V=-14$ mag as dividing luminosity according to \citeauthor{grebel01}'s dSph/dE distinction. The fit of the high luminosity part yields a CMR of 
   \begin{equation}
      (V-I)_{\mathrm{50}}=(-0.043\pm0.007)\cdot M_V+(0.370\pm0.105)
   \end{equation}
   and a RMS of $0.090$. The lower luminosity branch yields a CMR of 
   \begin{equation}
      (V-I)_{\mathrm{50}}=(-0.001\pm0.008)\cdot M_V+(0.927\pm0.089)
   \end{equation}
   with a RMS of $0.117$. The slope of the CMR at lower luminosities changes significantly and the scatter around the relation at lower luminosities increases only slightly. Within the given uncertainty of the slope, the luminosity does not show a dependency on the V-I colour in this range. This is illustrated by the almost vertical red line in Fig. \ref{fig:cmd2} for the dSph regime. The two solid lines indicate the different slopes of the relation.\\
   Our sample of background galaxies shows a trend towards redder colours at fainter magnitudes. For the understanding of this behaviour we use the \verb*#GALEV# stellar population models \citep{kotulla} for illustration purposes. \verb*#GALEV# provides the redshift evolution of the integrated light properties of an E-type galaxy (i.e. an exponentially declining star formation history with decay time of 1 Gyr) of $5\cdot10^{11}$M$_{\sun}$. The thick line in Fig. \ref{fig:cmd} shows the evolution of the position of this galaxy in the CMD with increasing redshift. The dark grey section of the line indicates the Virgo cluster redshift range, the red section of the line represents the evolution of the galaxy's position in the CMD with increasing redshift up to $z\approx0.65$. The progessively redder colours of background elliptical galaxies at fainter magnitudes is also shown by the \verb*#GALEV# model.

  \subsection{Scaling relations}
   \begin{figure}
   \centering
   \includegraphics[width=0.5\textwidth]{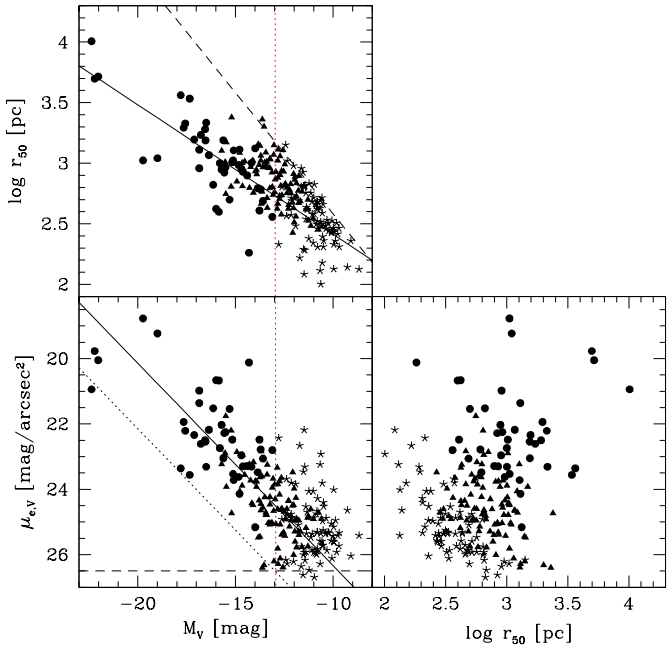}
      \caption{Scaling relations of all Virgo cluster objects (see text). Dashed lines: detection limit of $\mu_{e,V}=26.5$ mag/arcsec$^2$, solid lines: linear fit to all datapoints with $M_V<-14$ mag in the $r-M$ diagram (in $\mu-M$ diagram: $-18<M_V<-13$ mag), dotted black line: $2\sigma$ confidence interval, dotted red line: completeness limit of $M_V=-13.0$ mag, filled circles: spectroscopically confirmed Virgo cluster members, filled triangles: remaining objects with VCC index, asterisks: probable Virgo cluster members.}
         \label{fig:scaling}
   \end{figure}   
   The parameters which are used to distiguish dwarf galaxies from giant galaxies are luminosity and size. Hence, how both parameters scale to each other, or to other physical properties, is of particular interest. A change can be a hint for a different origin of the different types of galaxies - in this case giants, dEs and dSphs - or at least a hint for different evolution. In Fig. \ref{fig:scaling} we present the common three scaling relations, the luminosity-surface brightness diagram (bottom left panel), the luminosity-size diagram (top left panel) and the size-surface brightness diagram (bottom right panel). Shown are all 213 early-type galaxies which are considered to belong to the Virgo cluster.\\
   The dashed line in the two left panels indicates the surface brightness detection limit of $\mu_{e,V}\approx 26.5$ mag/arcsec$^2$. The solid line in the $r-M$ diagram represents the fit to all objects which are brighter than $M_V=-14$ mag. At that magnitude the slope of the CMR changed significantly. In the $\mu_e-M$ diagram the data is fitted to $M_V\leq-13$ for reasons which are discussed in Sect. \ref{sec:complete}. Furthermore, fitting the data down to that magnitude we are sure not to run into the completeness limit of our data.\\
   Regarding the luminosity-surface brightness diagram, the trend given by the Es and dEs ($M_V<-14$ mag) is obviously continued by the faint galaxies. The scatter around the trend does not seem to change significantly at fainter magnitudes. But the detection limit affects the completeness at fainter magnitudes. We see the same results in the luminosity-size diagram. The faint galaxies seem to continue the trend given by the bright galaxies and the scatter does not seem to increase. Also seen in this plot, the detection limit (dashed line) becomes important at lower luminosities. At faint magnitudes, we expect more galaxies, exhibiting surface brightnesses beyond the detection limit of $\mu_e\approx26.5$ mag. The size-surface brightness diagram does not reveal a trend. But it shows that the uncatalogued galaxies (asterisks) tend to have half-light radii $\lesssim10$ arcseconds, their mean $r_{50}$ $\approx4$ arcsec corresponds to 320 parsecs while they tend to have on average a lower surface brightness than the previously known member galaxies.

  \subsection{Completeness}\label{sec:complete}
   The crucial point in estimating a luminosity down to which the data is complete, is the scatter of the galaxies' parameters. We do not know whether the scatter remains constant at magnitudes fainter than our completeness limit. Hence, we do not know the full parameter distribution of galaxies, which is why we refrain from creating a sample of artificial galaxies based on that same parameter distribution.\\
   We use the $\mu_e-M$ relation to determine the completeness of our data. \citet{graham} used a correlation between the S\'{e}rsic index $n$ and luminosity to transform the linear $\mu_0-M$ relation into a curved $\mu_e-M$ relation. As a result, we cannot assume a linear trend over the entire luminosity range. We are not able to give a reliable curved trend since there are too few datapoints at high luminosities. We therefore investigate the trend given by all galaxies with luminosities $-18<M_V<-13$ mag. The lower luminosity is chosen to increase the confidence of the trend but not to run into incompleteness. Regarding \citeauthor{graham}'s Fig. 12 we consider it reliable to linearize the curve in our fitting interval. This trend is given in the $\mu_e-M$ plot of Fig. \ref{fig:scaling} and is not valid for higher luminosities, because there will be a turning point in the curved relation. At lower luminosites we do not expect large errors because of the asymptotic convergency to the linear $\mu_0-M$ relation. Our linear fit leads to
   \begin{equation}
      \mu_{e,V}=(0.62\pm0.09)\cdot M_V+(32.54\pm1.27).
   \end{equation}
   with a RMS of $=1.00$. Since we do not observe a change in the scaling relations we assume that the scatter around the relation also remains constant. We use the scatter around the relation to determine the completeness of our data sample. In particular, we use the $2\sigma$ confidence interval of the luminosity-surface brightness relation to determine the intersect with the detection limit of $26.5$ mag$\cdot$arcsec$^{-2}$. For all selected objects the scatter of that scaling relation is given by RMS, leading to the intersect at $M_V=-13.0$ mag. The $2\sigma$ interval defines a confidence range of 96\%. We expect to miss 2\% of the Virgo objects on the diffuse side at a luminosity of $M_V\approx-13.0$ mag. Thus, the data is rather complete down to that luminosity, given a constant scatter around the $\mu_e-M$ relation. 

  \subsection{Luminosity Function}
   \begin{figure}
   \centering
   \includegraphics[width=0.5\textwidth]{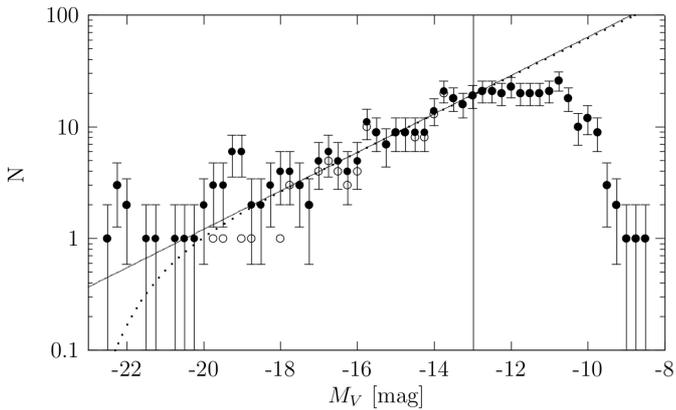}
      \caption{V-band luminosity function of the Virgo cluster sample (bin width: 0.5 mag, sampling step: 0.25 mag). Open symbols denote only early type galaxies, solid symbols denote datapoints which were corrected by SDSS data. The solid line indicates a linear fit performed in $-18.8\leq M_V \leq -13.0$ mag. The Schechter function fitted in the same interval with fixed $M^*_V=-21.8$ mag is represented by the dotted curve. The vertical line at $M_V=-13.0$ mag represents the assumed completeness limit. The errors come from Poissonian statistics.}
         \label{fig:lf}
   \end{figure}   
The LF of a cluster is defined as number of galaxies which are found within certain magnitude bins. To determine our V-band LF we choose a bin width of 0.5 mag and created sampling steps every 0.25 mag. Hence, the values are not independent from each other.\\
\citet{schechter} proposed an analytic approximation for the LF: 
   \begin{equation}
      N(x)dx=\Phi^*x^\alpha e^{-x}dx
   \end{equation}
  with 
   \begin{equation}
      x=10^{-0.4(M-M^*)}
   \end{equation}
Derivation of the faint end slope $\alpha$ of the Schechter function is critically dependent on the completeness of the dataset.\\
In order to get a reliable LF, we took also into account VCC galaxies
which were not investigated in this study. This concerns mainly
late-type galaxies, as well as early-type galaxies that were either
located close to a star or have been only partly covered. To get
information about the early-types, we used SDSS g- and r-band
magnitudes. For the late-type galaxies we took g- and r-band
photometry of Meyer et al. (in prep., following a similar procedure as
in \citealp{lisk07}) based on SDSS data. The g- and r-band magnitudes
of the additional 21 VCC galaxies, which are located within our
fields, were converted to V-band magnitudes using the transformations
from \citet{jester}. In Fig. \ref{fig:lf} we present the LF of the
V-band data. Solid symbols denote our data, open symbols represent
datapoints which were corrected by the additionally inserted
galaxies.\\ 
Due to the specifically selected target area of the Virgo core, encompassing all of the large ellipticals M84, M86 and M87, we refrain from attempting a fit of the characteristic magnitude (also see the discussion in Sect. \ref{sec:dis_lf}). Instead of using a Schechter function, we perform a linear fit in the interval $-18.8\leq M_V\leq-13.0$ mag. The bright end limit is chosen because of the jump in galaxy counts at $M_V = -19$ mag. It also corresponds to the typical division between "dwarf" and "normal" galaxies, and it is still safely away from typical values of $M^*$ in a Schechter luminosity function. The faint end limit is our completeness limit.
We use \verb+gnuplot+'s implementation of the nonlinear least-squares Marquardt-Levenberg algorithm to perfom an error weighted linear fit to the data. That has been done for three different fields and the whole data sample as displayed in Tab. \ref{table:slopes}. It turns out that faint galaxies seem more concentrated around M86/M84. For comparison, in the same fitting interval we also performed the Schechter function fit with fixed $M^*_V=-21.8$ mag, corresponding roughly to \citeauthor{sandage85}'s $M^*_B=-20.8$ mag. The results are also shown in Tab. \ref{table:slopes}. We find that the faint end slopes derived with a Schechter fit are slightly shallower in each case, but always lie within the errors. Fig. \ref{fig:lf} illustrates that they are almost indistinguishable from each other.\\
Given the areal variation of the faint end slope, we give the overall faint end slope as
   \begin{equation}
      \alpha=-1.50\pm0.05\pm0.12
   \end{equation}

\begin{table}
\caption{Faint end slopes.}             
\label{table:slopes}      
\centering                          
\begin{tabular}{c c c c c}        
\hline\hline                 
Field & Area & $n$ & $\alpha_{lin}$ & $\alpha_{Sch}$ \\    
\hline                        
M87 & $\left. \begin{array}{c}\textrm{RA}>187.3\deg \\ \textrm{DEC}<13.0\deg \end{array} \right.$ & 59 & $1.31\pm0.08$ & $1.29\pm0.06$ \\
M86 & $\left. \begin{array}{c}\textrm{RA}<186.9\deg \\ \textrm{DEC}>12.5\deg \end{array} \right.$ & 79 & $1.54\pm0.11$ & $1.47\pm0.08$ \\
between & else & 100 & $1.46\pm0.10$ & $1.41\pm0.07$ \\
\hline                                   
complete & all & 238 & $1.50\pm0.05$ & $1.43\pm0.04$ \\
\hline            
\end{tabular}\\
\tablefoot{$n$: number of galaxies located in an area, $\alpha_{lin}$: faint end slope of a linear fit, $\alpha_{Sch}$: faint end slope of a Schechter function. The errors are the fit's statistical errors. For comparison of the selected area see also Fig. \ref{fig:coord}.} \\
\end{table}
This faint end slope fits into the range of slopes determined by other
Virgo studies \citep{sandage85,trentham,rine08,saba03} and is also comparable to slopes determined for other nearby clusters \citep{hilk03,misgeld08,misg09,ferguson,beijersbergen,penn07,rine08}.\\

\section{Discussion}
\subsection{Sample selection}
The quality of the sample selection is best visible in the colour magnitude diagram in Fig. \ref{fig:cmd}.\\
The identification of background galaxies by adopting the S\'{e}rsic index larger than 1.5 as background criterion at apparent magnitudes fainter than $m_V\approx18$ mag excludes all objects with a compact appearance. Doing this, we would remove compact ellipticals\footnote{fainter than the known cEs but brighter than UCDs} from the Virgo cluster sample, if they were present. The probability of observing a faint background galaxy with a deVaucouleurs profile at faint magnitudes is much higher than observing a compact dwarf elliptical. This argumentation becomes stronger when regarding the CMD. The sample of objects excluded by their S\'{e}rsic index (open circles) continues the trend given by the redshift confirmed background objects (open triangles).\\
Another hint for dealing with background galaxies is the trend to extreme red colours of these galaxies at apparent magnitudes around $m_V=20$ mag. This trend can be explained by the colour change of ellipticals with increasing redshift, as shown by \verb+GALEV+ models (red line in the CMD)\footnote{The "turning point" to extreme red colours itself depends on the galaxy's mass.}.\\
Also excluded from our Virgo sample are ultra-compact dwarf galaxies (UCDs) since they are very small and did not pass our applied size criterion. When regarding the known UCDs in Virgo \citep{hase05,evst07}, we see that there are many other objects which appear similar to UCDs but are obviously unresolved background objects.\\
Due to the very red colours expected for background elliptical galaxies, we adopted an interval around the CMR inside which objects were considered Virgo cluster members. Of course, this criterion holds only when there is no change in the slope of the CMR at lower luminosities and the scatter remains Gaussian. While both of these conditions are not met (see Sect. \ref{sec:cmd}), we have accounted for this ($5\sigma$ interval) and only four objects were subsequently regarded as background objects.\\
As seen in Fig. \ref{fig:cmd}, however, there are velocity-confirmed background galaxies that have colours consistent with that of Virgo cluster galaxies. As a result, it is still possible that some of the objects in our sample of 77 probable galaxies are indeed background objects. That there are two objects with a large S\'{e}rsic index located in the Virgo member CMD region could be a hint for background objects. On the other hand, these two objects could also be compact Virgo ellipticals. One of them is shown in Fig. \ref{fig:crit12}. If they are the result of stripped spirals, they should have kept their bulge properties \citep{chil09}, including their red colour \citep{balc94}. Their colours ($V-I\simeq1$) are a bit bluer than M32 \citep[$V-I\simeq1.2$,][]{laue98}. A final statement of this issue can only be made when spectroscopic data of these galaxies are available. 

\subsection{Colour magnitude relation}

There is a change in the slope of the Virgo CMR at $M_V\approx-14$
mag: at fainter magnitudes, we observe no correlation between colour
and magnitude anymore. However, our sample might be biased at faint
magnitudes, since the completeness limit is $M_V=-13.0$ mag. In order
to assess whether the shallower depth of our I-band images could have
preferentially excluded galaxies with blue colours from our sample
just below the completeness limit (as we required detection in both
bands), we consider the scatter of our galaxies in colour and in surface
brightness just above the completeness limit. Due to the relatively
large spread in surface brightness and the fact that its \emph{faint end}
governs the completeness limit (Fig. \ref{fig:scaling}), the majority of galaxies around the
limit actually has sufficiently high S/N that no colour bias should
occur among them. Assuming the colour and surface brightness scatter
is similar below the limit, we estimate that down to $M_V=-11.7$ mag we should
not have lost more than 6 galaxies that could potentially be bluer
than average and thus affect the CMR slope. This number is consistent
with the 4 objects that were detected in the V-band only. Thus,
while the detection of the faintest targets may be biased towards red
colours, no significant bias should be present down to $M_V=-11.7$ mag, giving
confidence that the change of the CMR slope at $M_V\approx-14$ is
real.\\
In the E and dE luminosity range ($M_V\lesssim-14$ mag) we observe the
commonly known negative slope of the CMR. This is usually explained by
the larger potential wells of the more massive galaxies, which made it
easier for them to recycle their metal-enriched gas, leading to a higher
metallicity and thus redder colour. For the dSphs ($M_V>-14$ mag) no
such dependency is seen anymore. However, the intrinsic colour spread of the dSphs is not significantly larger than for the dEs\footnote{When not considering the uncertainty in the calibration zeropoint, which would apply equally to all galaxies, the RMS of our errors is 0.045 mag at these magnitudes, while the observed scatter is 0.117 mag, leading to an intrinsic scatter of 0.108 mag. The same number calculated for the dEs would be 0.090 mag.}.
This seems to be at odds with an extrapolation of the results of
\citet{gavazzi} for Virgo early-type galaxies: fainter galaxies show
a more extended star formation history, which should result in a
broader colour distribution. However, our data only contains galaxies
of Virgo's core region. It is likely that dwarf galaxies in this region
lost all their gas already at an early epoch of the cluster
evolution, due to tidal forces from the cluster potential and from
massive member galaxies, as well as ram pressure stripping from the
intracluster medium. It thus seems plausible that the faint, diffuse
dSphs in the Virgo core region have had a similar short period of star
formation, leading to similar red colours with small scatter.
We speculate that significant enrichment was only possible at
magnitudes brightward of the change in slope, which lead to the usual
mass-metallicity relation.

\subsection{Scaling relations}
Regarding the scaling relations of Fig. \ref{fig:scaling}, there are two results to report. First, dSphs seem to continue the trends given by their more luminous counterparts. And second, the detection limit of $\mu_{e,V}=26.5$ mag/arcsec$^{2}$ in connection with the scatter around the luminosity-surface brightness relation limits the completeness to $M_V\lesssim-13.0$ mag.\\
We excluded high surface brightness objects at lower luminosites from our sample, i.e. excluded all possible cEs and UCDs from our sample. 
This affects on the one hand the scatter of the found relations, on the other hand it could also affect the relations's shape itself. Only a dozen of Virgo UCDs are known, not affecting the overall numbers of member galaxies.\\
A change in the slope of the CMR is observed at $M_V\simeq-14$ mag. The data begins to become incomplete one magnitude fainter, but reaches $M_V\simeq-9$ mag. Hence, the dSphs which differ in their behaviour from the more luminous galaxies (dEs and Es) in the CMR are incomplete, almost over their whole luminosity range. This makes it impossible to prove whether the scaling trends given by the Es and dEs are indeed continued by the dSphs. We can only say: In the face of the scaling relations, the dSphs seem to be the extension of the dE population. In the light of the change in the CMR at $M_V\approx-14$ mag, this would lead to the argumentation that dSphs and dEs share the same origin (given by the physical parameters) but differ in their evolution (given by the chemical parameter colour).\\

\subsection{Luminosity function and completeness}\label{sec:dis_lf}
Due to observation of the Virgo core region, we are unable to independently determine the characteristic magnitude M* of the Virgo cluster. Schechter found that his function is a good fit to the LF of a whole cluster. We fit the Schechter function to a sample which is not representative for the Virgo cluster since our field were chosen such that the three giant ellipticals (M87 and M86/M84) are included. If we shift our field of view in any direction we would have only two or even one of them included, resulting in a more certain estimation of the turning point $M^*$. Because of that uncertain behaviour at the bright end of the LF we performed only a linear fit using a fitting interval which takes into account both the uncertainty at the bright end and at the faint end.\\
Our estimation of the completeness is rather conservative by using a $2\sigma$ confidence interval. Hence, within the fitting interval the data should be complete and therefore the faint end slope reliable.
The obtained faint end slope is $\alpha=-1.50\pm0.17$ for linear fit and $\alpha=-1.43\pm0.09$ for the Schechter function fit respectively. This is in good agreement with the investigation of \citet{saba03} who found a faint end slope of $\alpha =-1.4\pm0.2$ for the central $0.8\deg$ around M87, hence for the core region. Following
\citet{moore99} who use the Tully-Fisher relation to calculate the LF
of the Virgo cluster from the mass function of their simulations,
there is consistency between theory and observations. On the other
hand, simulations based on the favoured $\Lambda$CDM cosmological
model expect a faint end slope of the LF steeper than $-1.8$ as
deduced from the CDM mass function \citep{tully}. This observational
discrepancy to the theory is often observed. The favoured explanation
for the missed observed galaxies is supressed star formation in lowest
DM halos. Since we observe the core region of the cluster, disruption
should also play a role. Due to the gravitational potential of the
most massive galaxies -- located in the center of the cluster --
we can expect that the less massive galaxies are tidally distorted, some of them might have even been disrupted. This idea was already suggested by \citet{saba03} supported by their findings of $\alpha=-1.8$ in the outer regions of the Virgo cluster (distance to M87 $\ge 1.6\deg$). The disruption scenario is also supported by \citet{miho05} who find a "complex substructure of Virgo's diffuse intracluster light". They discuss in particular four streamers which they relate to ongoing tidal stripping of dwarf galaxies. The tidally distorted galaxies will eventually get stripped by the gravitational
 potential of their massive counterparts. In this process, the compact nuclei of nucleated dwarf ellipticals might end up as UCDs. Since the fraction of remnant nuclei among the UCD population is unknown -- many of them might be just giant globular clusters -- we did not consider them in our study. In any case, the progenitor galaxies of remnant nuclei-UCDs in general are in the luminosity range $-19<M_V<-15$, thus missing some of them does not affect the very faint end of the luminosity function.
Also stripped spiral galaxies -- ending up as compact ellipticals -- are not considered due to their high S\'{e}rsic index. Even if there might be undiscovered cEs, their number should be
small, since they may originate from giant spiral galaxies. And
simulations show, that initially large dark matter halos -- containing a giant spiral galaxy which is stripped during its evolution -- are
rare. Hence, cEs should not contribute tremendously to the faint end
of the LF neither. Finally, totally disrupted galaxies can not be
observed. The question arises what fraction of partially disrupted/transformed galaxies has been missed by applying our selection criteria.\\
Due to selection criteria 1,2,8, and 9 of Sect. \ref{sec:sample} we reject mostly spectroscopically known galaxies. We corrected the LF for the rejected VCC members, but we also reject two uncatalogued galaxies.  They might have magnitudes brighter than $M_V=-13$, and therefore, would have had a weak impact on the LF faint end slope. All galaxies rejected by criteria 6 and 7 are candidates for Virgo cluster members, but due to their faintness lie beyond the completeness limit considered for the determination of
the faint end slope of the LF.\\ 
Furthermore we also miss very low surface brightness objects which are
too diffuse to be detected. Due to the applied confidence interval of
the completeness determination we assume that we miss only one of
these objects ($1\%$). Also the assumption of a constant scatter around the $\mu_e-M$ relation to determine the completeness can be called into question. On the other hand, the faint end slope does not change significantly ($\alpha=-1.42$) when setting the detection limit to $M_V=-13.6$ mag.\\
Besides observational biases, also theoretical reasons may explain the lack of
dwarf galaxies. Local feedback effects \citep{dekel} or the
reionization of the Universe \citep{klypin} could inhibit the collapse
of gas into small haloes, leading to DM halos with no visible
counterpart. Additionally, \citeauthor{klypin} offer a plausible
expanation of the missing satellites for a Milky Way-like halo. They argue that high velocity
clouds (HVCs) are numerous enough when extrapolating their local
(Milky Way) abundance to a larger volume. Whether a dark matter halo
hosts a dwarf galaxy or an HVC depends primarily on the circular
velocity of the halo. But it is at least questionable if one can simply exend this argumentation to a galaxy cluster with its different evolution.


\section{Summary}
CFHT V- and I-band data of Virgo cluster's core region, taken in 1999 with the CFH12K instrument, have been analysed. Applying morphological criteria, 295 galaxies were chosen for photometric investigation. Knowing the redshift of the luminous galaxies, applying a surface brightness criterion at lower luminosities and regarding the location of these galaxies in the CMD, 216 of these galaxies are considered to belong to the Virgo galaxy cluster. 64 of these objects have previously been unknown. The detection limit of the image data is determined at $\mu_{e,V}=26.5$ mag/arcsec$^2$ for the faintest objects at $M_V=-8.7$ mag. In the dSph magnitude regime the detection limit may bias the sample towards redder galaxies.\\
The CMD of the Virgo cluster sample reveals a change in the slope of
the early-type CMR at $M_V\approx-14$ mag -- the distincting
luminosity for dEs and dSphs. Provided that the change is real, we
conclude that this luminosity corresponds to a specific galaxy mass which
is needed to retain its gas. At higher masses, this leads to an
extended star formation period and to self-enrichment of
the gas, resulting in higher
metallicities, and thus in redder colours. We do not observe a
significantly increasing scatter around the CMR of dSphs, which we
interpret as a comparably short star formation for both types of galaxies, the dEs and the dSphs. The fact that all dSphs have the same $V-I$ colour can be interpreted as a similar metallicity for this galaxy type. Since the cluster's core region is observed, we conclude
that the SFH of these faint galaxies was truncated at a very early
epoch of the cluster evolution, resulting in similar $V-I$ colours.\\ 
The investigation of the scaling relations (luminosity-size-surface
brightness) shows no difference between dEs and dSphs. 
In the face of this continuing trend, while a definitive conclusion is hampered by the completeness limit of $M_V=-13.0$ mag, dEs and dSphs seem consistent with having the same origin.

\begin{acknowledgements}
We thank Hagen Meyer for providing his unpublished photometric measurements for late-type galaxies in our region and Patrick C\^ot\'e for his comparison of our uncatalogued objects to the Next Generation Virgo cluster Survey (NGVS) data. He confirmed that all our uncatalogued Virgo objects have a matching object in the NGVS data. Special thanks are adressed to Mischa Schirmer for his support on THELI issues with the 'old' CFH12K data.\\
S.L.\ is supported by the ESO Studentship Programme. S.L.\ and T.L.\ are supported within the framework of the Excellence Initiative by the German Research Foundation (DFG) through the Heidelberg Graduate School of Fundamental Physics (grant number GSC 129/1). This research has made use of the NASA/IPAC Extragalactic Database (NED) which is operated by the Jet Propulsion Laboratory, California Institute of Technology, under contract with the National Aeronautics and Space Administration. 
\end{acknowledgements}

\bibliographystyle{aa} 

\Online
\begin{appendix}
\section{Tables}

\onllongtab{3}{
\begin{longtable}{cccccccccc}
\caption{Properties of VCC uncatalogued Virgo cluster objects found in this study but not included in the VCC}\\
\multicolumn{10}{l}{{\textbf{ID:} galaxy identification number in our catalogue sorted by $\alpha$}}\\
\multicolumn{10}{l}{{\textbf{$\alpha$ (J2000):} Right Ascension}}\\
\multicolumn{10}{l}{{\textbf{$\delta$ (J2000):} Declination}}\\
\multicolumn{10}{l}{{\textbf{$M_V$:} absolute V-band magnitude in Virgo cluster distance (adopted $m-M=31.09$ mag)}}\\
\multicolumn{10}{l}{{\textbf{$M_I$:} absolute I-band magnitude in Virgo cluster distance (adopted $m-M=31.09$ mag)}}\\
\multicolumn{10}{l}{{\textbf{$(V-I)_{50}$:} colour $V-I$ computed by the flux enclosed by an ellipse with half-light radius $r_{50}$}}\\
\multicolumn{10}{l}{{\textbf{$\epsilon$:} ellipticity at half-light aperture}}\\
\multicolumn{10}{l}{{\textbf{$r_{50}$:} half-light radius determined by the half-light semi-major axis $a_{50}$ and ellipticity $\epsilon$ of an ellipse by $r_{50}=a_{50}\sqrt{1-\epsilon}$}}\\
\multicolumn{10}{l}{{\textbf{$\mu_{e,V}$:} effective surface brightness in V-Band determined by the ellipse enclosing half of the total flux}}\\
\multicolumn{10}{l}{{\textbf{Reference:} literature reference catalogue if object was previously found}}\vspace{0.5cm}\\
\hline
\hline
ID & $\alpha$ (J2000) & $\delta$ (J2000) & $M_V$ & $M_I$ & $(V-I)_{50}$ & $\epsilon$ & $r_{50}$ & $\mu_{e,V}$ & Reference\\
 & [deg] & [deg] & [mag] & [mag]& [mag] & & [pc] & [mag/arcsec$^2$] &  \\ 
\hline\\
\endfirsthead
\caption{Continued.} \\
\hline
ID & $\alpha$ (J2000) & $\delta$ (J2000) & $M_V$ & $M_I$ & $(V-I)_{50}$ & $\epsilon$ & $r_{50}$ & $\mu_{e,V}$ & Reference\\
 & [deg] & [deg] & [mag] & [mag]& [mag] & & [pc] & [mag/arcsec$^2$] & \\ 
\hline\\
\endhead
\hline
\endfoot
\endlastfoot

$1$ & $186.08182$ & $12.32702$ & $-11.64 \pm 0.04$ & $-12.77 \pm 0.04$ & $1.016 \pm 0.055$ & $0.20$ & $641$ & $25.75$ & \\
$2$ & $186.09055$ & $12.60981$ & $-10.58 \pm 0.07$ & $-11.85 \pm 0.07$ & $0.967 \pm 0.085$ & $0.60$ & $518$ & $25.59$ & \\
$3$ & $186.17314$ & $12.51108$ & $-11.52 \pm 0.02$ & $-12.37 \pm 0.02$ & $0.854 \pm 0.020$ & $0.05$ & $195$ & $23.46$ & \\
$4$ & $186.18954$ & $12.77586$ & $-9.27 \pm 0.13$ & $-11.08 \pm 0.13$ & $0.995 \pm 0.094$ & $0.30$ & $138$ & $24.64$ & a\\
$5$ & $186.25246$ & $13.04010$ & $-11.33 \pm 0.01$ & $-12.02 \pm 0.01$ & $0.855 \pm 0.024$ & $0.45$ & $414$ & $24.70$ & a\\
$6$ & $186.30690$ & $13.07214$ & $-10.56 \pm 0.04$ & $-11.82 \pm 0.04$ & $1.000 \pm 0.053$ & $0.20$ & $387$ & $25.73$ & a\\
$7$ & $186.32645$ & $12.40963$ & $-10.22 \pm 0.06$ & $-10.93 \pm 0.06$ & $0.819 \pm 0.086$ & $0.20$ & $276$ & $25.34$ & \\
$8$ & $186.33395$ & $13.14852$ & $-11.63 \pm 0.01$ & $-12.61 \pm 0.01$ & $0.922 \pm 0.044$ & $0.30$ & $904$ & $26.36$ & a\\
$9$ & $186.35361$ & $13.10954$ & $-10.06 \pm 0.03$ & $-10.85 \pm 0.03$ & $0.858 \pm 0.080$ & $0.05$ & $335$ & $26.11$ & a\\
$10$ & $186.36690$ & $12.33431$ & $-11.14 \pm 0.04$ & $-12.23 \pm 0.04$ & $1.079 \pm 0.060$ & $0.50$ & $566$ & $25.47$ & \\
$11$ & $186.41136$ & $12.82371$ & $-11.25 \pm 0.01$ & $-11.96 \pm 0.01$ & $0.949 \pm 0.019$ & $0.15$ & $239$ & $24.06$ & \\
$12$ & $186.43758$ & $12.78776$ & $-10.67 \pm 0.01$ & $-11.74 \pm 0.01$ & $1.055 \pm 0.017$ & $0.20$ & $130$ & $23.24$ & \\
$13$ & $186.44778$ & $12.42193$ & $-12.39 \pm 0.02$ & $-13.24 \pm 0.02$ & $0.959 \pm 0.014$ & $0.27$ & $367$ & $23.68$ & \\
$14$ & $186.50537$ & $13.22742$ & $-11.07 \pm 0.02$ & $-11.46 \pm 0.02$ & $0.808 \pm 0.036$ & $0.05$ & $379$ & $25.36$ & \\
$15$ & $186.54506$ & $12.48158$ & $-12.12 \pm 0.01$ & $-13.27 \pm 0.01$ & $0.963 \pm 0.018$ & $0.05$ & $490$ & $24.86$ & \\
$16$ & $186.57158$ & $12.83289$ & $-10.52 \pm 0.02$ & $-11.64 \pm 0.02$ & $0.981 \pm 0.041$ & $0.25$ & $314$ & $25.25$ & a\\
$17$ & $186.60013$ & $12.41685$ & $-8.68 \pm 0.09$ & $-10.61 \pm 0.09$ & $1.135 \pm 0.111$ & $0.10$ & $133$ & $25.41$ & \\
$18$ & $186.60918$ & $12.65277$ & $-10.26 \pm 0.06$ & $-11.74 \pm 0.06$ & $1.103 \pm 0.070$ & $0.40$ & $321$ & $25.31$ & \\
$19$ & $186.61797$ & $12.97582$ & $-10.63 \pm 0.02$ & $-11.58 \pm 0.02$ & $1.010 \pm 0.014$ & $0.05$ & $100$ & $22.92$ & \\
$20$ & $186.65869$ & $11.89181$ & $-11.39 \pm 0.02$ & $-12.01 \pm 0.02$ & $0.908 \pm 0.053$ & $0.05$ & $568$ & $25.92$ & \\
$21$ & $186.68047$ & $12.29557$ & $-12.62 \pm 0.01$ & $-13.45 \pm 0.01$ & $0.909 \pm 0.016$ & $0.05$ & $633$ & $24.92$ & \\
$22$ & $186.68597$ & $13.18775$ & $-9.71 \pm 0.07$ & $-10.03 \pm 0.07$ & $0.833 \pm 0.115$ & $0.35$ & $291$ & $25.73$ & a\\
$23$ & $186.70673$ & $12.17863$ & $-10.86 \pm 0.02$ & $-11.81 \pm 0.02$ & $0.916 \pm 0.033$ & $0.35$ & $314$ & $24.75$ & \\
$24$ & $186.76381$ & $13.28985$ & $-10.49 \pm 0.03$ & $-11.17 \pm 0.03$ & $0.706 \pm 0.046$ & $0.30$ & $229$ & $24.52$ & \\
$25$ & $186.76797$ & $13.45586$ & $-9.88 \pm 0.05$ & $-10.77 \pm 0.05$ & $0.967 \pm 0.075$ & $0.40$ & $228$ & $24.95$ & \\
$26$ & $186.80318$ & $13.22085$ & $-10.39 \pm 0.04$ & $-11.35 \pm 0.04$ & $0.923 \pm 0.070$ & $0.20$ & $303$ & $25.36$ & \\
$27$ & $186.81429$ & $13.41314$ & $-11.24 \pm 0.05$ & $-12.23 \pm 0.05$ & $1.153 \pm 0.080$ & $0.05$ & $681$ & $26.47$ & \\
$28$ & $186.81993$ & $12.53548$ & $-9.98 \pm 0.03$ & $-10.77 \pm 0.03$ & $0.769 \pm 0.059$ & $0.10$ & $214$ & $25.15$ & \\
$29$ & $186.83177$ & $12.22115$ & $-11.44 \pm 0.02$ & $-11.84 \pm 0.02$ & $0.756 \pm 0.049$ & $0.05$ & $653$ & $26.17$ & \\
$30$ & $186.83179$ & $13.08700$ & $-10.81 \pm 0.04$ & $-12.02 \pm 0.04$ & $0.896 \pm 0.058$ & $0.05$ & $332$ & $25.32$ & a\\
$31$ & $186.83453$ & $11.69554$ & $-10.99 \pm 0.03$ & $-11.75 \pm 0.03$ & $0.886 \pm 0.063$ & $0.20$ & $520$ & $25.93$ & \\
$32$ & $186.83725$ & $12.57477$ & $-11.71 \pm 0.01$ & $-12.82 \pm 0.01$ & $1.211 \pm 0.012$ & $0.25$ & $165$ & $22.66$ & \\
$33$ & $186.87318$ & $11.73457$ & $-10.36 \pm 0.02$ & $-10.99 \pm 0.02$ & $0.727 \pm 0.052$ & $0.45$ & $288$ & $24.88$ & \\
$34$ & $186.89905$ & $12.62434$ & $-12.43 \pm 0.02$ & $-12.87 \pm 0.02$ & $0.792 \pm 0.040$ & $0.50$ & $1416$ & $26.16$ & c\\
$35$ & $186.93100$ & $11.96804$ & $-11.52 \pm 0.02$ & $-12.58 \pm 0.02$ & $1.029 \pm 0.049$ & $0.10$ & $643$ & $26.00$ & \\
$36$ & $186.93483$ & $12.55751$ & $-10.67 \pm 0.03$ & $-11.59 \pm 0.03$ & $0.833 \pm 0.057$ & $0.10$ & $495$ & $26.28$ & \\
$37$ & $186.95143$ & $13.07543$ & $-12.79 \pm 0.01$ & $-13.57 \pm 0.01$ & $0.812 \pm 0.007$ & $0.02$ & $214$ & $22.43$ & \\
$38$ & $186.97136$ & $12.38317$ & $-10.81 \pm 0.03$ & $-11.57 \pm 0.03$ & $0.830 \pm 0.053$ & $0.05$ & $417$ & $25.82$ & \\
$39$ & $187.02469$ & $12.83764$ & $-11.02 \pm 0.02$ & $-11.89 \pm 0.02$ & $0.922 \pm 0.028$ & $0.18$ & $219$ & $24.06$ & \\
$40$ & $187.03290$ & $12.40231$ & $-9.96 \pm 0.03$ & $-10.69 \pm 0.03$ & $0.802 \pm 0.070$ & $0.15$ & $271$ & $25.61$ & \\
$41$ & $187.06409$ & $12.56028$ & $-10.10 \pm 0.01$ & $-11.27 \pm 0.01$ & $0.979 \pm 0.034$ & $0.45$ & $275$ & $25.05$ & d\\
$42$ & $187.07405$ & $11.84793$ & $-10.85 \pm 0.03$ & $-12.36 \pm 0.03$ & $1.295 \pm 0.064$ & $0.05$ & $457$ & $25.99$ & \\
$43$ & $187.08340$ & $12.81626$ & $-10.56 \pm 0.02$ & $-11.40 \pm 0.02$ & $0.922 \pm 0.031$ & $0.10$ & $150$ & $23.81$ & \\
$44$ & $187.08376$ & $13.36015$ & $-10.97 \pm 0.06$ & $-11.54 \pm 0.06$ & $0.852 \pm 0.095$ & $0.05$ & $669$ & $26.69$ & \\
$45$ & $187.12372$ & $11.97202$ & $-10.54 \pm 0.03$ & $-11.54 \pm 0.03$ & $0.802 \pm 0.050$ & $0.05$ & $248$ & $24.97$ & \\
$46$ & $187.13344$ & $12.98780$ & $-9.83 \pm 0.05$ & $-10.74 \pm 0.05$ & $0.952 \pm 0.112$ & $0.30$ & $321$ & $25.92$ & \\
$47$ & $187.16617$ & $12.97799$ & $-11.48 \pm 0.03$ & $-12.08 \pm 0.03$ & $0.901 \pm 0.053$ & $0.20$ & $650$ & $25.93$ & \\
$48$ & $187.18599$ & $11.99365$ & $-10.50 \pm 0.02$ & $-11.34 \pm 0.02$ & $0.848 \pm 0.030$ & $0.20$ & $204$ & $24.40$ & \\
$49$ & $187.19551$ & $12.64202$ & $-11.09 \pm 0.03$ & $-12.28 \pm 0.03$ & $0.979 \pm 0.038$ & $0.05$ & $480$ & $25.85$ & \\
$50$ & $187.20824$ & $12.79630$ & $-11.92 \pm 0.01$ & $-12.80 \pm 0.01$ & $0.974 \pm 0.012$ & $0.20$ & $303$ & $23.84$ & \\
$51$ & $187.21622$ & $12.79762$ & $-11.45 \pm 0.01$ & $-12.22 \pm 0.01$ & $0.895 \pm 0.014$ & $0.20$ & $191$ & $23.31$ & \\
$52$ & $187.22379$ & $13.19753$ & $-10.00 \pm 0.04$ & $-10.98 \pm 0.04$ & $1.028 \pm 0.093$ & $0.30$ & $279$ & $25.43$ & \\
$53$ & $187.24644$ & $12.04175$ & $-10.81 \pm 0.03$ & $-12.12 \pm 0.03$ & $0.968 \pm 0.068$ & $0.25$ & $481$ & $25.88$ & \\
$54$ & $187.24792$ & $11.92320$ & $-11.54 \pm 0.03$ & $-12.35 \pm 0.03$ & $0.889 \pm 0.088$ & $0.05$ & $841$ & $26.62$ & \\
$55$ & $187.28853$ & $12.49628$ & $-10.52 \pm 0.04$ & $-12.16 \pm 0.04$ & $1.112 \pm 0.057$ & $0.45$ & $456$ & $25.71$ & \\
$56$ & $187.33977$ & $12.46774$ & $-10.10 \pm 0.04$ & $-11.19 \pm 0.04$ & $1.021 \pm 0.066$ & $0.06$ & $268$ & $25.57$ & e\\
$57$ & $187.38072$ & $12.57001$ & $-12.15 \pm 0.01$ & $-13.15 \pm 0.01$ & $1.021 \pm 0.011$ & $0.37$ & $373$ & $23.80$ & \\
$58$ & $187.39923$ & $12.75197$ & $-9.97 \pm 0.03$ & $-10.94 \pm 0.03$ & $0.879 \pm 0.034$ & $0.20$ & $134$ & $24.01$ & \\
$59$ & $187.42303$ & $12.49928$ & $-9.99 \pm 0.04$ & $-10.95 \pm 0.04$ & $0.958 \pm 0.069$ & $0.26$ & $291$ & $25.59$ & \\
$60$ & $187.42992$ & $12.65521$ & $-9.69 \pm 0.04$ & $-10.76 \pm 0.04$ & $0.976 \pm 0.069$ & $0.10$ & $204$ & $25.33$ & \\
$61$ & $187.44878$ & $12.57161$ & $-9.10 \pm 0.06$ & $-9.62 \pm 0.06$ & $0.876 \pm 0.126$ & $0.20$ & $229$ & $26.04$ & \\
$62$ & $187.47418$ & $12.62167$ & $-12.03 \pm 0.02$ & $-12.33 \pm 0.02$ & $0.670 \pm 0.035$ & $0.20$ & $952$ & $26.21$ & c\\
$63$ & $187.50749$ & $12.94782$ & $-10.81 \pm 0.03$ & $-11.42 \pm 0.03$ & $0.766 \pm 0.054$ & $0.45$ & $432$ & $25.31$ & \\
$64$ & $187.51823$ & $12.50990$ & $-9.76 \pm 0.11$ & $-10.44 \pm 0.11$ & $0.786 \pm 0.158$ & $0.37$ & $260$ & $25.41$ & \\
$65$ & $187.52582$ & $12.68849$ & $-12.15 \pm 0.01$ & $-12.90 \pm 0.01$ & $0.904 \pm 0.017$ & $0.20$ & $516$ & $24.76$ & \\
$66$ & $187.53278$ & $12.38887$ & $-10.80 \pm 0.02$ & $-12.15 \pm 0.02$ & $0.989 \pm 0.027$ & $0.05$ & $203$ & $24.28$ & \\
$67$ & $187.59959$ & $12.43587$ & $-10.91 \pm 0.04$ & $-12.40 \pm 0.04$ & $1.084 \pm 0.053$ & $0.05$ & $463$ & $25.95$ & b\\
$68$ & $187.61784$ & $12.98260$ & $-11.37 \pm 0.02$ & $-12.58 \pm 0.02$ & $1.148 \pm 0.048$ & $0.05$ & $432$ & $25.35$ & \\
$69$ & $187.62703$ & $13.09423$ & $-11.40 \pm 0.02$ & $-12.52 \pm 0.02$ & $0.985 \pm 0.028$ & $0.28$ & $346$ & $24.53$ & \\
$70$ & $187.63330$ & $12.38538$ & $-11.48 \pm 0.01$ & $-12.67 \pm 0.01$ & $1.256 \pm 0.009$ & $0.27$ & $121$ & $22.18$ & \\
$71$ & $187.63419$ & $12.86409$ & $-10.68 \pm 0.03$ & $-11.63 \pm 0.03$ & $1.087 \pm 0.060$ & $0.35$ & $424$ & $25.58$ & \\
$72$ & $187.64633$ & $13.18908$ & $-10.98 \pm 0.02$ & $-11.97 \pm 0.02$ & $0.979 \pm 0.041$ & $0.45$ & $382$ & $24.87$ & \\
$73$ & $187.66840$ & $12.62156$ & $-10.32 \pm 0.03$ & $-10.90 \pm 0.03$ & $0.854 \pm 0.059$ & $0.05$ & $340$ & $25.87$ & \\
$74$ & $187.70131$ & $12.33112$ & $-12.36 \pm 0.02$ & $-13.49 \pm 0.02$ & $1.155 \pm 0.035$ & $0.05$ & $1046$ & $26.27$ & \\
$75$ & $187.76324$ & $12.35285$ & $-11.78 \pm 0.01$ & $-12.48 \pm 0.01$ & $0.971 \pm 0.022$ & $0.15$ & $490$ & $25.09$ & \\
$76$ & $187.77147$ & $12.49412$ & $-10.64 \pm 0.02$ & $-11.18 \pm 0.02$ & $0.713 \pm 0.045$ & $0.15$ & $326$ & $25.35$ & \\
$77$ & $188.01996$ & $12.39497$ & $-10.83 \pm 0.02$ & $-11.88 \pm 0.02$ & $1.089 \pm 0.032$ & $0.30$ & $288$ & $24.67$ & \\
\hline \vspace{0.5cm}\\
\multicolumn{10}{l}{{$^a$: previously discovered by \citet{tully}}}\\
\multicolumn{10}{l}{{$^b$: previously discovered by \citet{trentham}}}\\
\multicolumn{10}{l}{{$^c$: previously discovered by \citet{impey}}}\\
\multicolumn{10}{l}{{$^d$: previously discovered by \citet{durrell}}}\\
\multicolumn{10}{l}{{$^e$: previously discovered by \citet{durr97}}}\\
\label{table:appendix}
\end{longtable}
}

\onllongtab{4}{
\begin{longtable}{cccccccccc}
\caption{Properties of VCC catalogued galaxies and spectroscopically confirmed cluster members investigated in this study}\\
\multicolumn{10}{l}{{\textbf{VCC index:} Index in the Virgo cluster catalogue \citep{binggeli}}}\\
\multicolumn{10}{l}{{\textbf{$z$:} redshift taken from NED}}\\
\multicolumn{10}{l}{{\textbf{remaining labels:} see Tab. \ref{table:appendix}}}\\ \vspace{0.5cm}\\
\hline
\hline
VCC & $z$ & $\alpha$ (J2000) & $\delta$ (J2000) & $M_V$ & $M_I$ & $(V-I)_{50}$ & $\epsilon$ & $r_{50}$ & $\mu_{e,V}$ \\
index & & [deg] & [deg] & [mag] & [mag]& [mag] & & [pc] & [mag/arcsec$^2$]\\  
\hline\\
\endfirsthead
\caption{Continued.} \\
\hline
VCC & $z$ & $\alpha$ (J2000) & $\delta$ (J2000) & $M_V$ & $M_I$ & $(V-I)_{50}$ & $\epsilon$ & $r_{50}$ & $\mu_{e,V}$ \\
index & & [deg] & [deg] & [mag] & [mag]& [mag] & & [pc] & [mag/arcsec$^2$]\\  
\hline\\
\endhead
\hline
\endfoot
\endlastfoot

$659$ & --- & $185.91063$ & $12.62773$ & $-12.84 \pm 0.01$ & $-13.81 \pm 0.01$ & $0.867 \pm 0.014$ & $0.20$ & $463$ & $23.84$ \\
$678$ & --- & $185.97064$ & $12.77302$ & $-13.13 \pm 0.02$ & $-14.58 \pm 0.02$ & $0.987 \pm 0.023$ & $0.05$ & $838$ & $25.02$ \\
$753$ & --- & $186.21513$ & $13.11117$ & $-15.47 \pm 0.00$ & $-16.20 \pm 0.00$ & $0.915 \pm 0.003$ & $0.05$ & $1485$ & $23.92$ \\
$765$ & --- & $186.26450$ & $13.24465$ & $-15.50 \pm 0.00$ & $-16.23 \pm 0.00$ & $0.933 \pm 0.002$ & $0.05$ & $557$ & $21.76$ \\
$736$ & $0.0035$ & $186.26556$ & $12.88692$ & $-22.02 \pm 0.00$ & $-23.19 \pm 0.00$ & $1.217 \pm 0.000$ & $0.08$ & $5188$ & $20.05$ \\
$767$ & --- & $186.27000$ & $13.07546$ & $-11.29 \pm 0.02$ & $-12.73 \pm 0.02$ & $0.923 \pm 0.038$ & $0.05$ & $499$ & $25.74$ \\
$775$ & --- & $186.29047$ & $12.38251$ & $-13.24 \pm 0.02$ & $-14.74 \pm 0.02$ & $1.029 \pm 0.019$ & $0.35$ & $1093$ & $25.08$ \\
$779$ & --- & $186.30466$ & $13.02545$ & $-14.16 \pm 0.00$ & $-15.22 \pm 0.00$ & $0.937 \pm 0.006$ & $0.05$ & $1027$ & $24.43$ \\
$781$ & $-0.0006$ & $186.31334$ & $12.71468$ & $-16.85 \pm 0.00$ & $-17.80 \pm 0.00$ & $0.946 \pm 0.002$ & $0.40$ & $908$ & $20.98$ \\
$793$ & $0.0063$ & $186.34026$ & $13.07070$ & $-14.62 \pm 0.00$ & $-15.23 \pm 0.00$ & $0.627 \pm 0.003$ & $0.25$ & $850$ & $23.30$ \\
$800$ & --- & $186.36095$ & $12.67696$ & $-13.40 \pm 0.01$ & $-14.03 \pm 0.01$ & $0.820 \pm 0.015$ & $0.35$ & $1029$ & $24.78$ \\
--- & $0.0081$ & $186.36891$ & $12.63667$ & $-15.09 \pm 0.01$ & $-15.82 \pm 0.01$ & $0.900 \pm 0.005$ & $0.25$ & $1276$ & $23.72$ \\
$803$ & --- & $186.37001$ & $12.49356$ & $-12.47 \pm 0.02$ & $-12.96 \pm 0.02$ & $0.678 \pm 0.031$ & $0.45$ & $979$ & $25.42$ \\
$804$ & --- & $186.37692$ & $12.97707$ & $-12.02 \pm 0.01$ & $-12.98 \pm 0.01$ & $1.031 \pm 0.033$ & $0.10$ & $815$ & $26.01$ \\
$810$ & --- & $186.38969$ & $13.22728$ & $-14.93 \pm 0.00$ & $-15.95 \pm 0.00$ & $0.969 \pm 0.002$ & $0.05$ & $707$ & $22.86$ \\
$814$ & --- & $186.40282$ & $12.84969$ & $-13.75 \pm 0.01$ & $-14.35 \pm 0.01$ & $0.793 \pm 0.011$ & $0.10$ & $1392$ & $25.45$ \\
$815$ & --- & $186.40504$ & $13.14373$ & $-15.70 \pm 0.00$ & $-16.80 \pm 0.00$ & $0.923 \pm 0.002$ & $0.15$ & $1271$ & $23.23$ \\
--- & $0.0019$ & $186.42220$ & $13.04777$ & $-13.59 \pm 0.00$ & $-14.18 \pm 0.00$ & $0.791 \pm 0.004$ & $0.30$ & $487$ & $23.06$ \\
$828$ & $0.0016$ & $186.42368$ & $12.81051$ & $-18.99 \pm 0.00$ & $-20.22 \pm 0.00$ & $1.227 \pm 0.000$ & $0.41$ & $1099$ & $19.23$ \\
$833$ & --- & $186.43604$ & $13.02213$ & $-14.28 \pm 0.00$ & $-15.61 \pm 0.00$ & $1.090 \pm 0.003$ & $0.05$ & $626$ & $23.23$ \\
$838$ & --- & $186.44617$ & $12.76034$ & $-14.08 \pm 0.01$ & $-14.94 \pm 0.01$ & $0.926 \pm 0.006$ & $0.10$ & $569$ & $23.17$ \\
$844$ & --- & $186.45151$ & $13.12248$ & $-12.37 \pm 0.01$ & $-13.29 \pm 0.01$ & $0.904 \pm 0.014$ & $0.30$ & $567$ & $24.60$ \\
$843$ & --- & $186.45448$ & $12.80439$ & $-12.74 \pm 0.01$ & $-13.86 \pm 0.01$ & $1.009 \pm 0.011$ & $0.18$ & $509$ & $24.16$ \\
$846$ & --- & $186.46049$ & $13.19765$ & $-15.53 \pm 0.00$ & $-16.58 \pm 0.00$ & $1.007 \pm 0.002$ & $0.19$ & $1023$ & $22.89$ \\
$850$ & --- & $186.46996$ & $13.19232$ & $-12.83 \pm 0.01$ & $-13.91 \pm 0.01$ & $0.924 \pm 0.011$ & $0.40$ & $761$ & $24.61$ \\
$854$ & $0.0023$ & $186.48209$ & $12.76975$ & $-14.18 \pm 0.01$ & $-15.04 \pm 0.01$ & $0.823 \pm 0.006$ & $0.64$ & $1005$ & $23.31$ \\
$871$ & $0.0048$ & $186.52353$ & $12.55965$ & $-16.50 \pm 0.00$ & $-17.48 \pm 0.00$ & $1.019 \pm 0.002$ & $0.34$ & $2159$ & $23.31$ \\
$872$ & --- & $186.52789$ & $12.86098$ & $-14.64 \pm 0.00$ & $-15.64 \pm 0.00$ & $1.075 \pm 0.002$ & $0.05$ & $649$ & $22.95$ \\
$876$ & --- & $186.54010$ & $12.39512$ & $-12.97 \pm 0.01$ & $-13.76 \pm 0.01$ & $0.891 \pm 0.009$ & $0.45$ & $734$ & $24.30$ \\
$881$ & $-0.0008$ & $186.54901$ & $12.94619$ & $-22.36 \pm 0.00$ & $-23.59 \pm 0.00$ & $1.439 \pm 0.000$ & $0.26$ & $10149$ & $20.94$ \\
$880$ & --- & $186.54947$ & $12.08699$ & $-11.02 \pm 0.02$ & $-12.66 \pm 0.02$ & $1.187 \pm 0.046$ & $0.30$ & $350$ & $24.91$ \\
$882$ & $0.0037$ & $186.56332$ & $12.96382$ & $-16.77 \pm 0.00$ & $-18.39 \pm 0.00$ & $1.071 \pm 0.001$ & $0.29$ & $1711$ & $22.61$ \\
$886$ & --- & $186.56369$ & $13.34086$ & $-10.75 \pm 0.03$ & $-11.48 \pm 0.03$ & $0.917 \pm 0.052$ & $0.30$ & $283$ & $24.71$ \\
$884$ & --- & $186.56528$ & $13.14302$ & $-13.54 \pm 0.01$ & $-14.39 \pm 0.01$ & $1.011 \pm 0.018$ & $0.20$ & $2005$ & $26.32$ \\
$892$ & --- & $186.58354$ & $12.51032$ & $-13.37 \pm 0.01$ & $-14.30 \pm 0.01$ & $0.900 \pm 0.005$ & $0.12$ & $429$ & $23.25$ \\
$896$ & --- & $186.59430$ & $12.78657$ & $-13.91 \pm 0.00$ & $-14.69 \pm 0.00$ & $0.926 \pm 0.005$ & $0.35$ & $812$ & $23.76$ \\
$898$ & --- & $186.59850$ & $13.37354$ & $-13.07 \pm 0.01$ & $-13.78 \pm 0.01$ & $0.955 \pm 0.011$ & $0.40$ & $550$ & $23.67$ \\
$903$ & --- & $186.61694$ & $12.92060$ & $-12.07 \pm 0.01$ & $-13.07 \pm 0.01$ & $0.877 \pm 0.009$ & $0.15$ & $267$ & $23.48$ \\
$916$ & $0.0043$ & $186.63835$ & $12.74302$ & $-15.99 \pm 0.00$ & $-17.29 \pm 0.00$ & $1.259 \pm 0.002$ & $0.05$ & $420$ & $20.66$ \\
$923$ & --- & $186.65141$ & $12.80280$ & $-13.32 \pm 0.01$ & $-14.08 \pm 0.01$ & $0.931 \pm 0.019$ & $0.05$ & $915$ & $25.03$ \\
$928$ & $-0.0008$ & $186.66585$ & $12.51358$ & $-15.71 \pm 0.00$ & $-16.70 \pm 0.00$ & $1.014 \pm 0.002$ & $0.42$ & $888$ & $22.03$ \\
$930$ & --- & $186.67143$ & $12.84534$ & $-13.24 \pm 0.01$ & $-14.40 \pm 0.01$ & $1.050 \pm 0.014$ & $0.47$ & $910$ & $24.46$ \\
$937$ & --- & $186.69421$ & $13.26675$ & $-12.71 \pm 0.03$ & $-13.61 \pm 0.03$ & $1.085 \pm 0.035$ & $0.05$ & $1139$ & $26.11$ \\
$940$ & $0.0047$ & $186.69611$ & $12.45402$ & $-17.11 \pm 0.00$ & $-18.14 \pm 0.00$ & $1.120 \pm 0.002$ & $0.10$ & $1566$ & $22.34$ \\
$941$ & --- & $186.69948$ & $13.37916$ & $-12.86 \pm 0.01$ & $-13.79 \pm 0.01$ & $0.988 \pm 0.009$ & $0.38$ & $421$ & $23.34$ \\
$942$ & --- & $186.70206$ & $12.39992$ & $-12.08 \pm 0.01$ & $-12.88 \pm 0.01$ & $0.881 \pm 0.020$ & $0.30$ & $472$ & $24.50$ \\
$951$ & $0.0069$ & $186.72653$ & $11.66373$ & $-17.57 \pm 0.00$ & $-18.79 \pm 0.00$ & $1.017 \pm 0.001$ & $0.34$ & $2129$ & $22.21$ \\
$956$ & --- & $186.73503$ & $12.96155$ & $-12.44 \pm 0.02$ & $-14.60 \pm 0.02$ & $1.333 \pm 0.025$ & $0.30$ & $678$ & $24.92$ \\
$959$ & --- & $186.74010$ & $12.42105$ & $-11.95 \pm 0.02$ & $-13.32 \pm 0.02$ & $1.101 \pm 0.035$ & $0.35$ & $770$ & $25.60$ \\
$962$ & --- & $186.74625$ & $12.50578$ & $-15.20 \pm 0.00$ & $-16.22 \pm 0.00$ & $1.018 \pm 0.006$ & $0.40$ & $2388$ & $24.73$ \\
$965$ & $0.0028$ & $186.76282$ & $12.56079$ & $-16.55 \pm 0.00$ & $-17.64 \pm 0.00$ & $1.024 \pm 0.002$ & $0.58$ & $1910$ & $22.50$ \\
$967$ & --- & $186.76572$ & $12.86657$ & $-12.99 \pm 0.01$ & $-13.98 \pm 0.01$ & $0.996 \pm 0.012$ & $0.40$ & $575$ & $23.85$ \\
$968$ & --- & $186.77528$ & $13.32370$ & $-13.11 \pm 0.01$ & $-13.98 \pm 0.01$ & $0.911 \pm 0.011$ & $0.40$ & $638$ & $23.95$ \\
$972$ & --- & $186.78506$ & $13.33580$ & $-14.78 \pm 0.01$ & $-15.89 \pm 0.01$ & $1.033 \pm 0.005$ & $0.15$ & $1218$ & $24.06$ \\
$977$ & --- & $186.79684$ & $12.03815$ & $-13.68 \pm 0.01$ & $-14.64 \pm 0.01$ & $1.017 \pm 0.005$ & $0.31$ & $462$ & $22.83$ \\
$978$ & --- & $186.79701$ & $12.11454$ & $-13.45 \pm 0.01$ & $-14.11 \pm 0.01$ & $0.795 \pm 0.010$ & $0.55$ & $1225$ & $24.71$ \\
$987$ & --- & $186.81487$ & $12.66054$ & $-13.62 \pm 0.01$ & $-14.04 \pm 0.01$ & $0.819 \pm 0.032$ & $0.30$ & $2303$ & $26.40$ \\
$996$ & --- & $186.83794$ & $13.11119$ & $-13.70 \pm 0.01$ & $-14.40 \pm 0.01$ & $0.919 \pm 0.011$ & $0.40$ & $1154$ & $24.64$ \\
$997$ & --- & $186.84235$ & $12.06869$ & $-13.74 \pm 0.01$ & $-14.70 \pm 0.01$ & $0.991 \pm 0.006$ & $0.15$ & $620$ & $23.63$ \\
$998$ & --- & $186.84776$ & $12.33168$ & $-13.79 \pm 0.01$ & $-14.23 \pm 0.01$ & $0.865 \pm 0.009$ & $0.30$ & $966$ & $24.34$ \\
$1004$ & --- & $186.85379$ & $13.40630$ & $-12.84 \pm 0.02$ & $-14.09 \pm 0.02$ & $1.068 \pm 0.033$ & $0.05$ & $1054$ & $25.81$ \\
$1008$ & --- & $186.86234$ & $11.94263$ & $-11.28 \pm 0.02$ & $-12.09 \pm 0.02$ & $0.835 \pm 0.033$ & $0.10$ & $336$ & $24.82$ \\
$1015$ & --- & $186.87306$ & $12.26924$ & $-11.30 \pm 0.02$ & $-12.18 \pm 0.02$ & $0.880 \pm 0.028$ & $0.38$ & $417$ & $24.87$ \\
$1014$ & --- & $186.87408$ & $12.25202$ & $-12.54 \pm 0.01$ & $-13.37 \pm 0.01$ & $0.847 \pm 0.020$ & $0.30$ & $853$ & $25.32$ \\
$1023$ & --- & $186.89363$ & $12.80369$ & $-12.57 \pm 0.02$ & $-13.50 \pm 0.02$ & $0.974 \pm 0.049$ & $0.15$ & $1283$ & $26.39$ \\
$1027$ & $0.0003$ & $186.91357$ & $12.87996$ & $-13.99 \pm 0.01$ & $-14.80 \pm 0.01$ & $0.937 \pm 0.013$ & $0.05$ & $1327$ & $25.16$ \\
$1037$ & --- & $186.92352$ & $12.48785$ & $-11.87 \pm 0.01$ & $-12.39 \pm 0.01$ & $0.775 \pm 0.022$ & $0.34$ & $532$ & $24.90$ \\
$1035$ & $-0.0017$ & $186.92545$ & $12.08967$ & $-15.30 \pm 0.00$ & $-16.24 \pm 0.00$ & $0.907 \pm 0.002$ & $0.20$ & $500$ & $21.54$ \\
$1041$ & --- & $186.94353$ & $11.74127$ & $-11.68 \pm 0.02$ & $-12.32 \pm 0.02$ & $0.886 \pm 0.034$ & $0.05$ & $486$ & $25.29$ \\
$1046$ & --- & $186.95621$ & $12.49964$ & $-11.59 \pm 0.02$ & $-12.64 \pm 0.02$ & $0.881 \pm 0.037$ & $0.05$ & $582$ & $25.77$ \\
$1051$ & --- & $186.97731$ & $12.60450$ & $-12.17 \pm 0.01$ & $-13.07 \pm 0.01$ & $0.936 \pm 0.026$ & $0.50$ & $674$ & $24.81$ \\
$1059$ & $0.0075$ & $187.00186$ & $11.94979$ & $-13.71 \pm 0.00$ & $-14.61 \pm 0.00$ & $0.946 \pm 0.004$ & $0.61$ & $608$ & $22.78$ \\
$1069$ & $0.0077$ & $187.02718$ & $12.89818$ & $-15.16 \pm 0.00$ & $-16.14 \pm 0.00$ & $1.016 \pm 0.003$ & $0.60$ & $1021$ & $22.48$ \\
$1070$ & --- & $187.02823$ & $12.97863$ & $-12.23 \pm 0.02$ & $-12.99 \pm 0.02$ & $0.802 \pm 0.027$ & $0.10$ & $601$ & $25.14$ \\
$1073$ & $0.0063$ & $187.03587$ & $12.09328$ & $-17.65 \pm 0.00$ & $-18.70 \pm 0.00$ & $1.127 \pm 0.001$ & $0.35$ & $1963$ & $21.94$ \\
$1077$ & --- & $187.04286$ & $12.80894$ & $-12.30 \pm 0.01$ & $-13.21 \pm 0.01$ & $0.930 \pm 0.018$ & $0.05$ & $421$ & $24.36$ \\
$1083$ & --- & $187.05096$ & $11.97039$ & $-12.02 \pm 0.01$ & $-13.07 \pm 0.01$ & $0.949 \pm 0.019$ & $0.20$ & $426$ & $24.48$ \\
$1081$ & --- & $187.05338$ & $13.01494$ & $-12.35 \pm 0.01$ & $-13.23 \pm 0.01$ & $0.958 \pm 0.015$ & $0.35$ & $629$ & $24.77$ \\
$1093$ & $0.0049$ & $187.07803$ & $11.70027$ & $-14.84 \pm 0.00$ & $-15.93 \pm 0.00$ & $0.998 \pm 0.004$ & $0.05$ & $965$ & $23.62$ \\
$1101$ & $0.0059$ & $187.09848$ & $13.19574$ & $-15.62 \pm 0.00$ & $-16.82 \pm 0.00$ & $1.034 \pm 0.003$ & $0.55$ & $1547$ & $23.05$ \\
$1103$ & --- & $187.10948$ & $12.34587$ & $-12.82 \pm 0.01$ & $-12.74 \pm 0.01$ & $0.520 \pm 0.026$ & $0.05$ & $1295$ & $26.28$ \\
$1104$ & $0.0057$ & $187.11690$ & $12.82368$ & $-16.36 \pm 0.00$ & $-17.36 \pm 0.00$ & $1.088 \pm 0.002$ & $0.30$ & $1164$ & $22.18$ \\
$1115$ & --- & $187.13548$ & $11.74473$ & $-14.11 \pm 0.00$ & $-14.94 \pm 0.00$ & $0.928 \pm 0.007$ & $0.10$ & $1052$ & $24.48$ \\
$1122$ & $0.0015$ & $187.17380$ & $12.91592$ & $-16.85 \pm 0.00$ & $-17.90 \pm 0.00$ & $1.100 \pm 0.002$ & $0.58$ & $1292$ & $21.36$ \\
$1123$ & $0.0063$ & $187.17764$ & $12.54976$ & $-14.79 \pm 0.00$ & $-15.79 \pm 0.00$ & $1.022 \pm 0.004$ & $0.19$ & $1296$ & $24.14$ \\
$1129$ & $0.0000$ & $187.18709$ & $12.80956$ & $-13.86 \pm 0.00$ & $-14.85 \pm 0.00$ & $1.022 \pm 0.005$ & $0.18$ & $620$ & $23.48$ \\
$1131$ & --- & $187.19077$ & $12.02186$ & $-13.45 \pm 0.01$ & $-14.31 \pm 0.01$ & $0.920 \pm 0.009$ & $0.20$ & $950$ & $24.78$ \\
$1136$ & --- & $187.20448$ & $12.13161$ & $-13.23 \pm 0.01$ & $-14.02 \pm 0.01$ & $0.880 \pm 0.010$ & $0.36$ & $1073$ & $25.03$ \\
$1139$ & --- & $187.21362$ & $11.95757$ & $-11.39 \pm 0.02$ & $-12.11 \pm 0.02$ & $0.767 \pm 0.029$ & $0.10$ & $417$ & $25.18$ \\
$1143$ & --- & $187.23154$ & $12.70682$ & $-12.62 \pm 0.01$ & $-13.55 \pm 0.01$ & $1.011 \pm 0.014$ & $0.05$ & $544$ & $24.59$ \\
$1147$ & --- & $187.24036$ & $11.95570$ & $-11.14 \pm 0.01$ & $-12.07 \pm 0.01$ & $0.942 \pm 0.025$ & $0.21$ & $304$ & $24.61$ \\
$1148$ & $0.0047$ & $187.24223$ & $12.66174$ & $-15.85 \pm 0.00$ & $-17.16 \pm 0.00$ & $1.280 \pm 0.002$ & $0.05$ & $396$ & $20.67$ \\
$1149$ & --- & $187.24571$ & $12.90790$ & $-13.84 \pm 0.01$ & $-15.09 \pm 0.01$ & $1.085 \pm 0.015$ & $0.05$ & $1432$ & $25.48$ \\
$1153$ & --- & $187.24922$ & $12.64835$ & $-13.99 \pm 0.00$ & $-14.97 \pm 0.00$ & $1.052 \pm 0.005$ & $0.38$ & $858$ & $23.75$ \\
$1157$ & --- & $187.25827$ & $12.43486$ & $-12.64 \pm 0.01$ & $-13.54 \pm 0.01$ & $0.949 \pm 0.019$ & $0.20$ & $863$ & $25.39$ \\
$1162$ & --- & $187.27148$ & $12.15374$ & $-11.87 \pm 0.02$ & $-12.73 \pm 0.02$ & $0.965 \pm 0.052$ & $0.10$ & $684$ & $25.78$ \\
$1161$ & --- & $187.27260$ & $12.03124$ & $-12.51 \pm 0.01$ & $-13.50 \pm 0.01$ & $0.962 \pm 0.023$ & $0.35$ & $655$ & $24.69$ \\
$1173$ & $0.0080$ & $187.31190$ & $12.97797$ & $-15.51 \pm 0.00$ & $-16.57 \pm 0.00$ & $1.088 \pm 0.002$ & $0.45$ & $921$ & $22.25$ \\
$1177$ & --- & $187.33018$ & $12.37704$ & $-13.04 \pm 0.01$ & $-14.19 \pm 0.01$ & $0.983 \pm 0.012$ & $0.68$ & $1173$ & $24.66$ \\
$1185$ & $0.0017$ & $187.34798$ & $12.45080$ & $-15.80 \pm 0.00$ & $-16.93 \pm 0.00$ & $1.165 \pm 0.002$ & $0.05$ & $1002$ & $22.74$ \\
$1191$ & --- & $187.36948$ & $12.49619$ & $-13.93 \pm 0.01$ & $-14.89 \pm 0.01$ & $0.965 \pm 0.006$ & $0.35$ & $882$ & $23.92$ \\
--- & $0.0041$ & $187.39005$ & $13.19570$ & $-13.11 \pm 0.00$ & $-13.90 \pm 0.00$ & $0.675 \pm 0.007$ & $0.35$ & $360$ & $22.80$ \\
$1202$ & --- & $187.39815$ & $13.21121$ & $-10.60 \pm 0.03$ & $-11.37 \pm 0.03$ & $0.879 \pm 0.053$ & $0.35$ & $298$ & $24.90$ \\
$1213$ & $0.0037$ & $187.41348$ & $12.54826$ & $-15.12 \pm 0.00$ & $-16.14 \pm 0.00$ & $1.041 \pm 0.004$ & $0.05$ & $1056$ & $23.53$ \\
$1216$ & --- & $187.42232$ & $12.04649$ & $-12.71 \pm 0.01$ & $-13.54 \pm 0.01$ & $0.864 \pm 0.026$ & $0.25$ & $797$ & $25.08$ \\
$1219$ & --- & $187.43367$ & $12.80547$ & $-13.38 \pm 0.01$ & $-14.25 \pm 0.01$ & $0.936 \pm 0.008$ & $0.19$ & $516$ & $23.55$ \\
$1229$ & --- & $187.44662$ & $13.07623$ & $-12.00 \pm 0.01$ & $-13.07 \pm 0.01$ & $0.993 \pm 0.017$ & $0.15$ & $325$ & $23.98$ \\
$1244$ & --- & $187.48473$ & $13.22007$ & $-13.24 \pm 0.01$ & $-13.77 \pm 0.01$ & $0.827 \pm 0.011$ & $0.08$ & $621$ & $24.23$ \\
$1251$ & --- & $187.50479$ & $13.11810$ & $-11.76 \pm 0.02$ & $-12.77 \pm 0.02$ & $0.983 \pm 0.030$ & $0.20$ & $471$ & $24.95$ \\
$1259$ & --- & $187.52544$ & $12.37726$ & $-13.50 \pm 0.01$ & $-14.51 \pm 0.01$ & $0.919 \pm 0.008$ & $0.55$ & $1011$ & $24.24$ \\
$1264$ & --- & $187.54538$ & $12.19552$ & $-14.66 \pm 0.00$ & $-15.80 \pm 0.00$ & $0.964 \pm 0.007$ & $0.05$ & $1004$ & $23.89$ \\
$1271$ & --- & $187.56360$ & $12.51589$ & $-12.16 \pm 0.01$ & $-13.16 \pm 0.01$ & $0.975 \pm 0.018$ & $0.20$ & $565$ & $24.95$ \\
$1279$ & $0.0045$ & $187.57256$ & $12.32848$ & $-19.73 \pm 0.01$ & $-20.92 \pm 0.01$ & $1.253 \pm 0.005$ & $0.17$ & $1054$ & $18.77$ \\
$1278$ & --- & $187.57257$ & $12.24105$ & $-12.93 \pm 0.01$ & $-13.91 \pm 0.01$ & $0.898 \pm 0.022$ & $0.20$ & $846$ & $25.06$ \\
$1277$ & --- & $187.57497$ & $12.04175$ & $-11.79 \pm 0.02$ & $-12.86 \pm 0.02$ & $1.035 \pm 0.038$ & $0.30$ & $593$ & $25.28$ \\
$1282$ & --- & $187.57590$ & $12.57139$ & $-11.75 \pm 0.02$ & $-12.55 \pm 0.02$ & $0.832 \pm 0.027$ & $0.57$ & $785$ & $25.40$ \\
$1286$ & --- & $187.60255$ & $12.79299$ & $-11.97 \pm 0.02$ & $-13.33 \pm 0.02$ & $1.093 \pm 0.029$ & $0.45$ & $729$ & $25.29$ \\
$1298$ & --- & $187.63908$ & $12.90059$ & $-13.70 \pm 0.01$ & $-14.53 \pm 0.01$ & $0.961 \pm 0.009$ & $0.40$ & $1051$ & $24.45$ \\
$1300$ & --- & $187.64432$ & $12.45821$ & $-12.65 \pm 0.01$ & $-13.71 \pm 0.01$ & $1.000 \pm 0.009$ & $0.15$ & $378$ & $23.65$ \\
--- & $0.0043$ & $187.69296$ & $12.09916$ & $-13.77 \pm 0.00$ & $-14.62 \pm 0.00$ & $0.729 \pm 0.006$ & $0.30$ & $406$ & $22.48$ \\
$1310$ & --- & $187.69531$ & $13.21397$ & $-11.91 \pm 0.01$ & $-12.88 \pm 0.01$ & $1.012 \pm 0.020$ & $0.18$ & $363$ & $24.26$ \\
$1313$ & $0.0042$ & $187.70210$ & $12.04514$ & $-14.31 \pm 0.00$ & $-14.65 \pm 0.00$ & $0.195 \pm 0.004$ & $0.35$ & $182$ & $20.12$ \\
$1314$ & --- & $187.70425$ & $13.22386$ & $-14.53 \pm 0.00$ & $-15.53 \pm 0.00$ & $1.031 \pm 0.005$ & $0.41$ & $1036$ & $23.56$ \\
$1316$ & $0.0044$ & $187.70596$ & $12.39114$ & $-22.20 \pm 0.01$ & $-23.60 \pm 0.01$ & $1.354 \pm 0.005$ & $0.09$ & $4976$ & $19.77$ \\
$1317$ & --- & $187.71080$ & $12.73658$ & $-13.64 \pm 0.01$ & $-14.56 \pm 0.01$ & $0.977 \pm 0.005$ & $0.15$ & $584$ & $23.60$ \\
$1335$ & --- & $187.76375$ & $12.07797$ & $-11.97 \pm 0.02$ & $-12.88 \pm 0.02$ & $0.921 \pm 0.032$ & $0.20$ & $520$ & $24.96$ \\
$1348$ & $0.0065$ & $187.81554$ & $12.33178$ & $-16.14 \pm 0.00$ & $-17.35 \pm 0.00$ & $1.235 \pm 0.004$ & $0.04$ & $663$ & $21.52$ \\
$1353$ & --- & $187.83095$ & $12.73802$ & $-15.22 \pm 0.00$ & $-16.27 \pm 0.00$ & $1.037 \pm 0.004$ & $0.17$ & $639$ & $22.20$ \\
$1352$ & $0.0062$ & $187.83150$ & $12.61151$ & $-14.40 \pm 0.01$ & $-15.50 \pm 0.01$ & $1.127 \pm 0.008$ & $0.30$ & $791$ & $23.29$ \\
$1381$ & --- & $187.93323$ & $12.61242$ & $-12.68 \pm 0.02$ & $-13.70 \pm 0.02$ & $1.003 \pm 0.028$ & $0.27$ & $651$ & $24.63$ \\
$1386$ & $0.0043$ & $187.96394$ & $12.65699$ & $-16.53 \pm 0.00$ & $-17.51 \pm 0.00$ & $1.059 \pm 0.003$ & $0.34$ & $1542$ & $22.54$ \\
$1389$ & $0.0029$ & $187.96672$ & $12.48177$ & $-15.57 \pm 0.00$ & $-16.65 \pm 0.00$ & $1.103 \pm 0.005$ & $0.28$ & $836$ & $22.28$ \\
$1399$ & $0.0016$ & $188.00314$ & $12.62034$ & $-14.70 \pm 0.00$ & $-15.89 \pm 0.00$ & $1.016 \pm 0.006$ & $0.48$ & $900$ & $22.96$ \\
$1413$ & --- & $188.03178$ & $12.43426$ & $-13.60 \pm 0.02$ & $-14.66 \pm 0.02$ & $1.066 \pm 0.018$ & $0.22$ & $928$ & $24.56$ \\
$1418$ & --- & $188.04730$ & $12.50683$ & $-14.22 \pm 0.01$ & $-15.33 \pm 0.01$ & $1.062 \pm 0.010$ & $0.36$ & $930$ & $23.73$ \\
$1438$ & --- & $188.14555$ & $12.64104$ & $-13.03 \pm 0.03$ & $-13.73 \pm 0.03$ & $0.887 \pm 0.047$ & $0.10$ & $1411$ & $26.20$ \\
$1448$ & $0.0086$ & $188.17004$ & $12.77097$ & $-17.80 \pm 0.00$ & $-18.72 \pm 0.00$ & $0.988 \pm 0.002$ & $0.20$ & $3639$ & $23.36$ \\
$1466$ & --- & $188.23056$ & $12.63519$ & $-12.28 \pm 0.02$ & $-12.86 \pm 0.02$ & $0.875 \pm 0.029$ & $0.20$ & $453$ & $24.35$ \\
$1493$ & --- & $188.32085$ & $12.58180$ & $-12.74 \pm 0.02$ & $-13.60 \pm 0.02$ & $0.944 \pm 0.018$ & $0.50$ & $537$ & $23.75$ \\
\hline \vspace{0.5cm}\\
\multicolumn{10}{l}{{objects without VCC index are spectroscopically confirmed Virgo cluster members}}\\

\label{table:appendix2}
\end{longtable}
}
\end{appendix}
\end{document}